\documentclass[a4paper,11pt]{article}
\usepackage{jheppub} 
\usepackage{orcidlink}

\usepackage{subfigure}
\usepackage{mathtools}
\usepackage{amsmath}
\usepackage{amsfonts}
\usepackage{amssymb}
\usepackage{bbold}
\usepackage{bm}
\usepackage{ esint }
\usepackage{orcidlink}

\newcommand{\br}{{\bf r}}

\newcommand{\bz}{{\bf z}}

\newcommand{\bk}{{\bf k}}
\newcommand{\bK}{{\bf K}}
\newcommand{\bv}{{\bf v}}

\newcommand{\bp}{{\bf p}}

\newcommand{\exclude}[1]{{}}
\long\def\exclude#1{}

\title{How collective modes terminate in the dispersion relation of neutrino plasmas.}

\author[a,b]{Damiano F.\ G.\ Fiorillo \orcidlink{0000-0003-4927-9850}} 
\affiliation[a]{Gran Sasso Science Institute (GSSI), L’Aquila, Italy,
Viale Francesco Crispi 7, 67100 L'Aquila (AQ), Italia}
\affiliation[b]{Istituto Nazionale di Fisica Nucleare (INFN), Sezione di Napoli, Complesso Universitario di Monte Sant’Angelo, Via Cintia, 80126 Napoli, Italy}

\abstract{Neutrino plasmas support flavor waves, collective oscillations of flavor, that neutrinos can resonantly emit or absorb when they fulfill the Cherenkov condition. For massless neutrinos, this condition leads to singular features in the dispersion relation: Landau-damped modes, for example, disappear abruptly as their phase velocity approaches the speed of light. We investigate how these singularities change when nonzero neutrino masses introduce a spread in velocities. We classify singular behavior of the frequency as a function of wavevector into \textit{i)} degeneracy points, where two modes merge, and \textit{ii)} abrupt mode termination, when the frequency approaches a branch cut of the flavor dielectric function. For each class, we derive a local expansion and determine where such behavior can occur. Degeneracy points, frequently encountered in axially symmetric dispersion relations in the fast limit, are not generic. Nonzero neutrino masses lift the degeneracy along the symmetry axis; in three-dimensional wavevector space, there might still be degeneracy lines, not necessarily intersecting the symmetry axis. Abrupt mode termination, instead, persists generically, except for special energy distributions which are entire functions. As a byproduct of this study, we show that a collisionless plasma with a Lorentzian neutrino-frequency distribution is exactly equivalent to a monochromatic one with an effective collision rate. Approximating a general frequency distribution by a sum of Lorentzians therefore substantially simplifies the dispersion relation.
}

\begin{document}
\maketitle
\flushbottom

\section{Introduction}\label{sec:introduction}

Neutrinos at the large densities reached in the outflow of supernovae (SNe) and neutron star mergers (NSMs) behave quite unlike the conventional picture of ghostly, non-interacting particles. Neutrino--neutrino interactions cannot be neglected, and are described by the Standard Model (SM) Hamiltonian
\begin{equation}\label{eq:interaction_Hamiltonian}
    \mathcal{H}_{\rm int}=\frac{G_F}{\sqrt{2}}\sum_{\alpha,\beta}\overline{\nu}_\alpha \gamma^\mu \nu_\alpha \overline{\nu}_\beta\gamma_\mu \nu_\beta
\end{equation}
in terms of the Fermi constant $G_F$ and the left-handed neutrino field $\nu_\alpha$ with flavor $\alpha=e,\mu,\tau$. The coherent potential sourced by flavor-mixed neutrinos, described by the density $\langle \overline{\nu}_\alpha \gamma^\mu \nu_\beta\rangle$, can trigger flavor exchanges over spatial and temporal scales defined by
\begin{equation}
    \mu=\sqrt{2}G_F n_\nu,
\end{equation}
where $n_\nu$ is the neutrino number density. The phenomenon induced by these mutual interactions is broadly known as collective flavor conversion (CFC)~\cite{Tamborra:2020cul,Volpe:2023met,Johns:2025mlm,Raffelt:2025wty,Fiorillo:2026gzy}. 

The onset of collective conversion can be understood through the dispersion relation of small flavor perturbations~\cite{Izaguirre:2016gsx, Capozzi:2017gqd,Airen:2018nvp,Yi:2019hrp,Fiorillo:2024bzm,Fiorillo:2024dik,Fiorillo:2025zio}.
This dynamics may be pictured in terms of individual neutrino flavor oscillations in a self-consistent potential, but in practice the most illuminating viewpoint is to treat the weak field mediating flavor exchange as an emergent degree of freedom, evoking a linear response from neutrinos~\cite{Fiorillo:2024bzm,Fiorillo:2024uki,Fiorillo:2024pns,Fiorillo:2025ank}. This is the collisionless-plasma description of a dense neutrino gas~\cite{Fiorillo:2025zio}. Its collective excitations carry the difference in lepton number (DLN) between two flavors and are called flavomons $\psi$~\cite{Fiorillo:2025npi}. In a medium whose DLN is dominated by $\nu_e$, for example, flavomons are emitted by a small population of ``flipped'' neutrinos with the opposite DLN through $\overline{\nu}_e\to\overline{\nu}_\mu\psi$
 and $\nu_\mu\to\nu_e\psi$. The inverse processes absorb them. The imbalance between emission and absorption gives rise either to flavor instabilities or to Landau damping. The nonlinear outcome of this dynamics is outside the scope of this paper; we merely refer to a representative sample of recent studies, e.g. Refs.~\cite{Martin:2019gxb,Capozzi:2017gqd,Bhattacharyya:2020jpj,Richers:2021nbx,Wu:2021uvt,Nagakura:2022kic,Cornelius:2023eop,Richers:2022bkd,Zaizen:2022cik,Abbar:2023ltx,Goimil-Garcia:2026flm,Fiorillo:2026jgw,Fiorillo:2026att}, and to the above-mentioned reviews~\cite{Tamborra:2020cul,Volpe:2023met,Johns:2025mlm,Raffelt:2025wty,Fiorillo:2026gzy}. For the rest of this work, we only focus on the linear theory of flavomons.

 The dispersion relation of flavomons and their classical counterparts, flavor waves, relates their frequency $\Omega_\bK$ to their wavevector $\bK$. It is determined by the zeros of the ``flavor dielectric function'', which describes the response of the plasma to a weak field, in analogy with electromagnetic plasma waves. At each wavevector, there are multiple solutions, that we organize in branches, defined by the continuous evolution of a given solution as a function of wavevector. Such branches can often merge or terminate abruptly at specific wavevectors; we call these points singularities, since the dispersion relation $\Omega_\bK$ is non-analytic in their neighborhood. 

 These singularities may also matter for the propagation of flavor waves. Wavepackets drift with their group velocity~\cite{Fiorillo:2025ank}, motivating a geometrical-optics description in inhomogeneous environments advocated in Ref.~\cite{Johns:2025yxa}. The practical challenge was taken on in Refs.~\cite{Fiorillo:2025gkw,Fiorillo:2026tee}, which show that gradients in matter density exert a force on flavomons, changing their wavevector until it may eventually cross a singular point in the dispersion relation.
 Even in a homogeneous medium, the evolving neutrino distribution can bring modes to a degeneracy. As emphasized in Refs.~\cite{Johns:2025yxa,Kost:2026ckc}, such degeneracies can invalidate approximations based on independently evolving flavor-wave modes. Understanding the local structure of singularities appear seems a prerequisite to assess the validity of these approximations.

 \begin{figure*}
     \includegraphics[width=\textwidth]{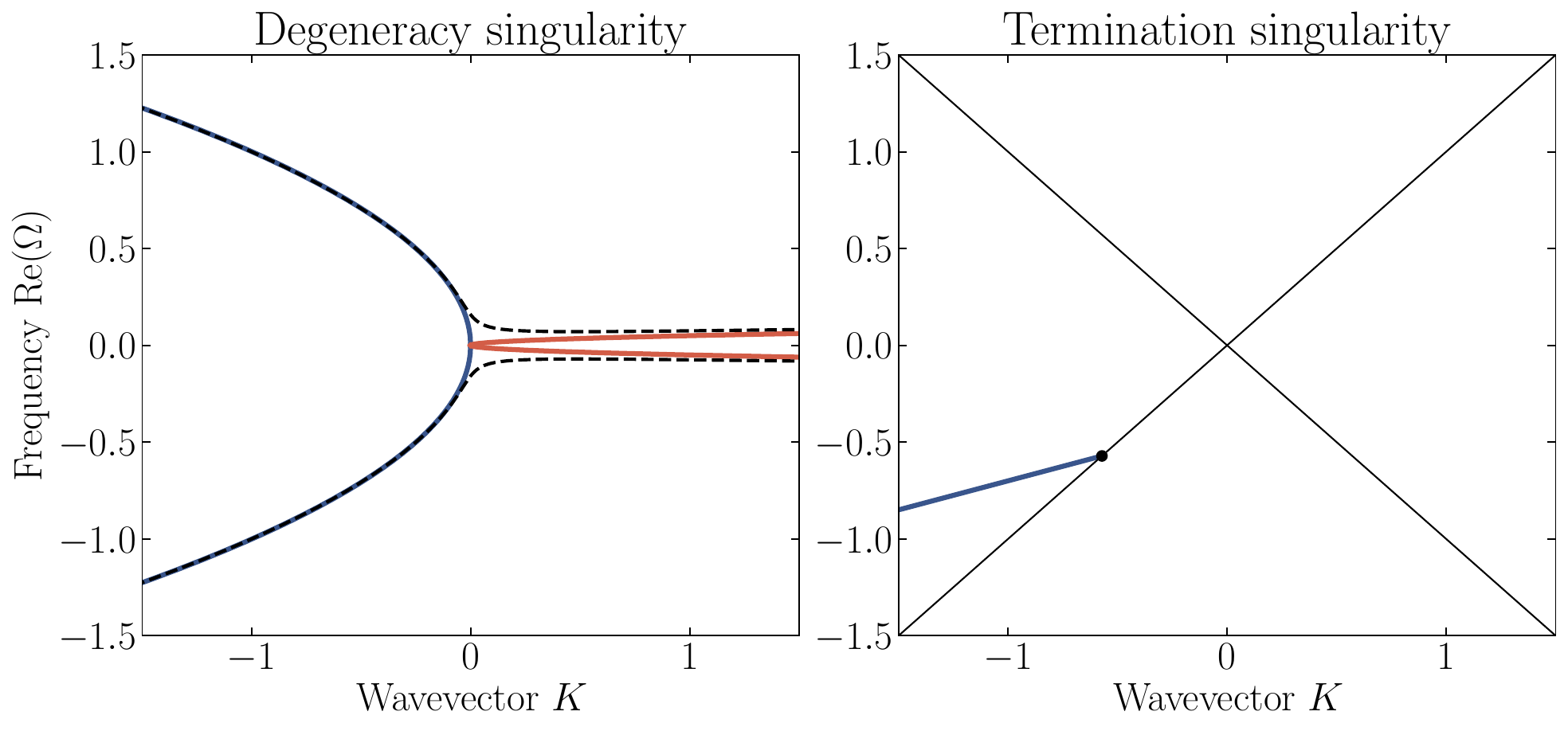}
     \caption{Schematic representation of degeneracy and termination singularities. A generic degeneracy singularity connects two pairs of complex branches, shown here through their real parts. In the fast limit, degeneracies outside the light cone have a pair of complex-conjugate branches on one side of the singularity. We show with a dashed curve the lifted degeneracy, which we find to be generically induced by nonzero neutrino masses. Termination singularities generically lie on the light cone, where the phase velocity equals the speed of light, in the fast limit; as we show in this work, nonzero vacuum masses shift their locations.}\label{fig:singularities}
 \end{figure*}

 We tackle this question here, relying on several recent theoretical developments: in particular, in the fast limit of vanishing neutrino masses, Refs.~\cite{Fiorillo:2024bzm,Fiorillo:2024uki} have shown that the singularities in the dispersion relation arise from two causes: \textit{i)} merging of multiple branches of flavor waves that become degenerate (\textit{degeneracy singularity}), and \textit{ii)} abrupt termination of damped modes when they touch a branch cut of the dispersion relation (\textit{termination singularity}). In the fast limit, the latter is very simple to identify; it coincides with the light cone, namely the surface where the phase velocity of the wave is equal to the speed of light. These two kinds of singularities are shown schematically in Fig.~\ref{fig:singularities}.

 When neutrino masses are included, however, the resonance condition is smeared by the spread in neutrino velocity, and the fate of singularities is not clear. A numerical solution of the dispersion relation is given in Refs.~\cite{Fiorillo:2024pns,Fiorillo:2025ank,Fiorillo:2025zio,Fiorillo:2025gkw,Fiorillo:2025kko}, which determined the analytical behavior in several asymptotic regimes, without however taking on the question of the singular points. Recently, Ref.~\cite{Kost:2026jrk} has importantly progressed by showing explicit examples where the singular points of the second type---abrupt termination of damped modes---disappear altogether for nonzero neutrino masses.

 We address this question through the complex analytic structure of the flavor dielectric function, deriving criteria for degeneracy and termination points and the local behavior of $\Omega_\bK$ near them. Termination points can survive nonzero neutrino masses; their absence in the examples of Ref.~\cite{Kost:2026jrk} is caused by the choice of Gaussian frequency distributions, which, as explained therein, are entire functions. Degeneracy points, however, are less ubiquitous than the fast limit suggests: nonzero vacuum frequencies generically lift them along the symmetry axis, while allowing degeneracy curves in three-dimensional $\bK$ space.
 
 As a byproduct of this study, we show that Lorentzian frequency distributions have the same linear response as monochromatic distributions with an effective collision rate, substantially simplifying physical insight and numerical calculations.

Sec.~\ref{sec:dispersion_relation} formulates the retarded dispersion relation. Sections~\ref{sec:degeneracies} and~\ref{sec:landau_damping} derive the criteria and local structure of degeneracy and termination singularities, respectively. In Sec.~\ref{sec:examples}, we solve the dispersion relation for two selected examples of slow (neutrino-mass-induced) instability and fast instability, exhibiting all the features we theoretically predict in the previous sections. Finally, we summarize our main results in Sec.~\ref{sec:summary} and discuss their possible implications.

\section{The dispersion relation of the neutrino plasma}\label{sec:dispersion_relation}

\subsection{Neutrino kinetic equations}

The dispersion relation of the plasma descends from its kinetic response to an external field. In practice, plasma kinetics is described by the density matrix $\rho_\bp$ for a neutrino mode with momentum $\bp$; in two flavors $\alpha=e,\mu$, this encodes the flavor state of neutrinos with momentum $\bp$ in the form
\begin{equation}
    \rho_\bp=\frac{1}{2}\begin{pmatrix}
        n_\bp+D_\bp && \psi^*_\bp\\
        \psi_\bp && n_\bp-D_\bp
    \end{pmatrix},
\end{equation}
where $n_\bp$ is the total occupation number, $D_\bp$ is the DLN, and $\psi_\bp$ measures the off-diagonal coherence, which is non-zero when neutrinos are in a flavor-mixed state. For antineutrinos, analogous quantities are denoted by an overbar $\overline{\rho}_\bp$; we adopt the flavor-isospin convention that antineutrinos carry the opposite DLN to neutrinos.

In the mean-field approximation, the kinetic equation for $\rho_\bp$ takes the form of a Vlasov equation~\cite{Dolgov:1980cq, Rudsky, Sigl:1993ctk,Vlasenko:2013fja,Serreau:2014cfa, Fiorillo:2024fnl, Fiorillo:2024wej,Fiorillo:2026byi}
\begin{equation}
    (\partial_t+\bv\cdot\partial_\br)\rho_\bp=i[\rho_\bp,{\sf\Omega}_{\bp,\rm vac}+{\sf\Omega}_{\bp,\rm mat}+{\sf\Omega_{\bp,\nu\nu}}].
\end{equation}
Here the vacuum energy splitting matrix ${\sf\Omega}_{\bp,\rm vac}=\mp w_{E} (c_V {\sf\sigma_3}-s_V {\sf\sigma_1})/2$ depends on the vacuum frequency $w_{E}=\delta m^2/2 E_\bp$, written in terms of the squared mass splitting $\delta m^2$ and the neutrino energy $E_\bp=|\bp|$. The vacuum mixing angle $\theta_V$ appears through $c_V=\cos2\theta_V$ and $s_V=\sin 2\theta_V$, and ${\sf\sigma}_i$ are the Pauli matrices. Normal ordering corresponds to $c_V>0$, while inverted ordering to $c_V<0$. In addition, electrons in matter source their own refractive potential ${\sf\Omega}_{\bp,\rm mat}=\lambda {\sf\sigma}_3/2$, with $\lambda=\sqrt{2}G_F n_e$ in terms of the electronic density, without including the non-relativistic flow velocity of matter; we neglect terms proportional to the identity which do not participate in flavor evolution. Finally, the neutrino--neutrino refractive potential inferred from Eq.~\ref{eq:interaction_Hamiltonian} is
\begin{equation}
    {\sf\Omega}_{\bp,\nu\nu}=\sqrt{2}G_F (\rho^\mu-\overline{\rho}^\mu) v_\mu,
\end{equation}
where
\begin{equation}
    \rho^\mu=\int \frac{d^3\bp}{(2\pi)^3}\rho_\bp v^\mu
\end{equation}
is the momentum-integrated density matrix, and similarly for antineutrinos.  Then the magnitude of the neutrino--neutrino interaction term $\sf\Omega_{\bp,\nu\nu}$ is measured by the energy scale $\mu=\sqrt{2}G_F n_\nu$. 

\subsection{Supernova-inspired hierarchy}

As a useful hierarchy to keep in mind, we remark that under typical conditions in the inner regions of a SN, the neutrino--neutrino interaction energy is typically of the order of
\begin{equation}
    \mu \simeq 0.6\,\mathrm{cm}^{-1}\, \frac{n_\nu}{10^{32}\,\mathrm{cm}^{-3}}.
\end{equation}
This should be compared with the neutrino vacuum splitting, which for the largest mass splitting is of the order of
\begin{equation}
    w_E\simeq 2\times 10^{-6}\,\mathrm{cm}^{-1}\, \frac{\delta m^2}{10^{-3}\, \mathrm{eV}^{2}}\, \frac{10\,\mathrm{MeV}}{E_\bp}.
\end{equation}
Finally, the matter refraction term 
\begin{equation}
    \lambda \simeq 60\,\mathrm{cm}^{-1}\,\frac{n_e}{10^{34}\,\mathrm{cm}^{-3}}.
\end{equation}
So a useful hierarchy to keep in mind, in the innermost regions of a SN, is $w_E\ll \mu\lesssim\lambda$. On the other hand, since in this work we focus on the analytical properties of the dispersion relation, we will not usually rely on this hierarchy.

\subsection{Linear response theory and dispersion relation}

The nonlinear nature of the kinetic equation makes it hard to obtain a general solution. However, we are interested here only in the much simpler problem of the behavior of small perturbations on top of an otherwise flavor-diagonal background. Non-flavor-diagonal backgrounds can also be slightly perturbed, if they correspond to an equilibrium, as has been recently proposed in Ref.~\cite{Johns:2025yxa}, although there is not yet any current practical scenario where this possibility appears. We stick to a flavor-diagonal background, and linearize the equations assuming that $\psi_\bp$ is very small, and that the background distributions $n_\bp$ and $D_\bp$ are unaffected by the evolution. This assumption becomes inconsistent in the case of an instability, but here we focus precisely on the properties of the linear modes. The linearized equations for $\psi_\bp$ then become
\begin{equation}\label{eq:eqs_motion}
    i(\partial_t + \bv \cdot \partial_\br)\psi_\bp=(w_E c_V-\lambda-\sqrt{2}G_F D\cdot v)\psi_\bp+\sqrt{2}G_F \psi\cdot v D_\bp+w_E s_V D_\bp.
\end{equation}
Here we have introduced the fields

\begin{equation}
    D^\mu=\int\frac{d^3\bp}{(2\pi)^3}(D_\bp-\overline{D}_\bp)v^\mu=\sum_\bp D_\bp v^\mu
\end{equation}
and $\psi^\mu=\sum_\bp \psi_\bp v^\mu$. Here we have introduced a notation that turns out to be particularly convenient, namely to define $\sum_\bp$ as the integral over phase space including antineutrinos as states of negative energies and opposite DLN, following the flavor-isospin convention.

The structure of the equation contains terms of three different nature on the right-hand side. The terms in parenthesis describe the frequency of evolution of each individual mode $\psi_\bp$. The second term $\sqrt{2}G_F \psi\cdot v D_\bp$ mixes all the different modes through the collective off-diagonal field $\psi^\mu$. The latter is precisely the weak potential field causing flavor mixing; the quanta of this field are called flavomons~\cite{Fiorillo:2025npi}. We note that the order of magnitude of the potential inducing flavor mixing is not merely $\mu$, since it is the DLN that enters the evolution. Therefore, we usually introduce a dimensionless parameter schematically measuring the ratio of the total DLN divided by the total number of neutrinos
\begin{equation}
    \epsilon\sim \frac{n_{\nu_e}-n_{\nu_\mu}-n_{\overline{\nu}_e}+n_{\overline{\nu}_\mu}}{n_\nu},
\end{equation}
such that the relevant interaction strength is measured by $\mu\epsilon$.
Finally, the last term $w_E s_V D_\bp$ is a genuinely flavor-violating term induced by mixing, which acts as a seed for the self-consistent modes of the system.

Since the equations are linear, we may look for the response $\psi_\bp$ for a monochromatic, plane field. This is usually done by introducing the normal modes and searching for solutions $\psi_\bp\propto e^{-iK\cdot X}$, where $K^\mu=(\Omega,\bK)$ is the four-vector of the wave and $X^\mu$ is the spatial four-vector. The seeding term $w_E s_V D_\bp$, which acts as an external sourcing term, can be neglected, since we are interested in the self-consistent modes of the system. The self-consistency condition then reads
\begin{equation}
    \psi_\bp=\frac{\sqrt{2}G_F D_\bp v^\mu}{K\cdot v + \sqrt{2}G_F D\cdot v + \lambda - w}\psi_\mu.
\end{equation}
We conveniently introduce the notation $w=w_E c_V$, and we assume throughout normal ordering $c_V>0$, so $w>0$ for neutrinos and $w<0$ for antineutrinos. After multiplying by $v^\mu$ and integrating over the momentum distribution, we obtain the condition
\begin{equation}
    \varepsilon^{\mu\nu}\psi_\nu=\left(g^{\mu\nu}-\chi^{\mu\nu}\right)\psi_\nu=0,
\end{equation}
where
\begin{equation}
    \chi^{\mu\nu}=\sum_\bp \frac{\sqrt{2}G_F D_{\bp}v^\mu v^\nu}{k\cdot v-w}
\end{equation}
is the flavor susceptibility, and $\varepsilon^{\mu\nu}$ is the flavor dielectric function. We have introduced here the shifted wavevector
\begin{equation}
    k^\mu=K^\mu + \sqrt{2}G_F D^\mu + \lambda u^\mu=(\omega,\bk),
\end{equation}
with $u^\mu=(1,0,0,0)$. If matter is moving, $u^\mu$ should be replaced with the four-component velocity of the flow.

While our derivation is based on identifying normal modes, namely exact solutions of the system which are oscillating with frequency $\Omega$, a similar result can be obtained by considering the response of the medium to an external forcing term. In that case, the asymptotic behavior at late times $t\to \infty$ can also be written as a superposition of modes, with the change that the susceptibility is modified
\begin{equation}\label{eq:susceptibility}
    \chi^{\mu\nu}=\sum_\bp \frac{\sqrt{2}G_F D_{\bp}v^\mu v^\nu}{k\cdot v-w+i0},
\end{equation}
where the $i0$ signifies that the integration contour should always be deformed in the complex plane as to pass below the singularity $k\cdot v - w=0$. (The asymptotic behavior also receives contributions from the branch cuts of the flavor susceptibility, which give power-law tails at late times~\cite{Fiorillo:2024bzm}.) The meaning of the deformation is quite intuitive physically; it corresponds to the fact that the response to a perturbation should only come after the perturbation itself appears, so one may insert the latter infinitely slowly by modifying the monochromatic perturbation $e^{-i\Omega t}\to \mathrm{lim}_{\eta\to +0}e^{-i\Omega t+\eta t}$. In practice, this means that the dispersion relation $\varepsilon^{\mu\nu}(\Omega,\bK)\psi_\nu=0$ should be continued from $\mathrm{Im}(\Omega)>0$ analytically. For this reason, we refer to it as \textit{retarded} dispersion relation, because it describes the retarded response of the medium after the perturbation has appeared.

This statement is the central premise of our work. It provides a definite and continuous prescription to obtain the modes determining the behavior of $\psi^\mu$. For $\mathrm{Im}(\Omega)>0$, the retarded dispersion relation and the normal-mode approach provide the same result. On the other hand, when $\mathrm{Im}(\Omega)<0$, the two approaches diverge. The normal-mode dispersion relation is purely real for real frequency, and therefore normal modes always appear in complex conjugate pairs. This also means that normal modes are singular at $\mathrm{Im}(\Omega)=0$, since they must pair with their conjugate partner giving rise to a non-differentiable behavior.

Instead, the retarded dispersion relation is not real. When $\mathrm{Im}(\Omega)$ passes from positive to negative, the integration contour should be deformed as to pass below the singularity in the complex plane. This endows the integral with an imaginary part associated with the part of the contour deformed in the complex plane.

This mathematical statement reflects a physical distinction between normal modes and retarded evolution. Normal modes describe perturbations with an exact exponential time dependence, whereas retarded evolution describes the response to a prescribed initial perturbation. Because neutrino momenta form a continuum, the discrete collective normal modes alone do not provide a complete description of this response. They must be supplemented by a continuum of singular eigenmodes with $\mathrm{Im}(\Omega)=0$, supported at the resonance $k\cdot v=w$. These Case–Van Kampen modes~\cite{VanKampen:1955wh,case1959plasma}, introduced in this context in Refs.~\cite{Fiorillo:2023mze,Fiorillo:2024bzm}, together with any discrete normal modes form a complete set of eigenmodes. A smooth initial perturbation excites a superposition of continuum modes, rather than an individual singular eigenmode. Their contributions can lose coherence through phase mixing, so that the collective amplitude $\psi^\mu$ decays even though each continuum mode has a real frequency. Thus, the collective evolution need not correspond to any single normal-mode frequency. Normal modes are particularly useful for identifying instabilities, but describing general perturbations requires the full continuum expansion; the retarded formulation captures this evolution more directly.

Instead, the retarded dispersion relation skips over this problem, by looking directly at the evolution of $\psi^\mu$ itself. For this reason, it includes new complex solutions with $\mathrm{Im}(\Omega)<0$ that arise from the deformed part of the contour in the negative complex half-plane. These solutions describe Landau damping~\cite{Landau:1946jc}, namely the exponential decrease in $\psi^\mu$ arising from the superposition of a continuum of Case-Van Kampen modes. For the same reason, the solutions of the retarded dispersion relation are perfectly regular when $\mathrm{Im}(\Omega)$ passes through zero.

In summary, we are interested in the physical modes that satisfy the retarded self-consistent condition $\varepsilon^{\mu\nu} \psi_\nu=0$, where $\varepsilon^{\mu\nu}$ is defined as the analytical continuation in Eq.~\ref{eq:susceptibility}. 
In practice, if we define the function
\begin{equation}
    \Phi(\omega,\bk)=\mathrm{det}\left[\varepsilon^{\mu\nu} g_{\mu\alpha} g_{\nu\beta}\right],
\end{equation}
then the existence of zero modes requires
\begin{equation}\label{eq:dispersion_relation_formal}
    \Phi(\omega_\bk,\bk)=0.
\end{equation}

This is the dispersion relation in its most compact and implicit form.  From now on, we will only use the shifted variables $\bk$ and $\omega_\bk$; the distinction with the physical frequency and wavevector matters in inhomogeneous and time-dependent conditions, but not for the task of determining the dispersion relation of flavor waves. Furthermore, we will use the notation $\omega_\bk$ whenever we refer to the on-shell frequency of flavomons, i.e.\ the solution of the implicit dispersion relation Eq.~\ref{eq:dispersion_relation_formal}, whereas we retain the notation $\omega$ whenever the frequency is treated as a free variable in the flavor dielectric function.

\subsection{Axial symmetry}

A particularly simple form is obtained under the often-made assumption of axially symmetric neutrino distribution, such that $D_\bp$ depends only on energy and on the cosine of the angle with the symmetry axis $v=\cos\theta$, and axially symmetric modes directed along $\bk=k_z \hat{\mathbf{z}}$. It is conventional then to introduce a simpler distribution defined in terms of $w=\delta m^2 c_V/2 E_\bp$ and $v$, such that
\begin{equation}\label{eq:axial_distribution}
    D_{v,w}=\frac{E_\bp^2}{4\pi^2}\left|\frac{dE_\bp}{dw}\right|D_\bp=\frac{(\delta m^2)^3 c_V^3}{32\pi^2 w^4}D_\bp.
\end{equation}
With this definition, all integrals $\sum_\bp$ should be replaced by $\sum_{v,w}=\int_{-1}^{+1} dv \int_{-\infty}^{+\infty} dw$; one can verify directly that $\sum_\bp D_\bp=\sum_{v,w} D_{v,w}$.

We will not use this specific case for our general results, but we will adopt it as a benchmark to exemplify these results. The flavor dielectric tensor in this case breaks down into separate pieces, since $\varepsilon_{za}$ and $\varepsilon_{0a}$ with $a=x,y$ vanish identically. Therefore, two degenerate modes arise immediately from the condition $\varepsilon_{xx}=\varepsilon_{yy}=0$, with $\psi^\mu=(0,1,0,0)$ and $\psi^\mu=(0,0,1,0)$ respectively. These modes are sometimes called axial-breaking, because their polarization vector $\psi^\mu$ breaks the axial symmetry, or transverse, since the spatial polarization vector is orthogonal to the symmetry axis. The explicit form of the dispersion relation, after integrating over the azimuthal angle, is simply
\begin{equation}
    1+\frac{\sqrt{2}G_F}{2}\sum_{v,w}\frac{D_{v,w}(1-v^2)}{\omega-k_z v-w+i0}=0.
\end{equation}
Introducing the family of integrals
\begin{equation}
    I_n=\int_{-1}^{+1}dv \int_{-\infty}^{+\infty} dw\frac{\sqrt{2} G_F D_{v,w}v^n}{\omega-k_z v-w+i0},
\end{equation}
the dispersion relation for the transverse modes is compactly
\begin{equation}
    I_0-I_2=-2.
\end{equation}

The modes polarized along the axis of symmetry, with $\psi^a=0$ for $a=x,y$, have a dispersion relation obtained from the determinant of the two-by-two block $zt$ in Eq.~\ref{eq:dispersion_relation_formal}
\begin{equation}\label{eq:dispersion_relation_longitudinal}
    (I_0-1)(1+I_2)-I_1^2=0.
\end{equation}

In the fast limit of vanishing neutrino masses, the dispersion relation simplifies even further, because $w$ disappears from the denominator, so that only the frequency-integrated angular distribution $D_v=\int_{-\infty}^{+\infty} dw D_{v,w}$ enters the integrals.

\section{Degeneracy singularities}\label{sec:degeneracies}

We now turn to the central goal of this paper, namely to identify the location of singularities in the behavior of $\omega_\bk$, the frequency of a given eigenmode as a function of $\bk$, and to classify their nature. The challenge is to obtain these singularities from the retarded dispersion relation. The first path to a singularity is that multiple solutions of $\Phi(\omega,\bk)=0$ merge. We focus on this first case of degeneracy singularities in this section.

\subsection{Superluminal modes in the fast limit}\label{sec:degeneracy_superluminal}

In the limit of vanishing neutrino masses $w\to 0$ (\textit{fast limit}), and superluminal flavomon modes $|\mathrm{Re}(\omega_\bk)|/|\bk|>1$, the possibility of multiple branches converging at a degeneracy point has been systematically discussed in Refs.~\cite{Yi:2019hrp,Fiorillo:2024uki}. We briefly review the arguments presented there. The reason this case is particularly simple is that the dispersion relation is purely real for $\omega_\bk$ real. For simplicity, we will simply say that the dispersion relation is real-symmetric
\begin{equation}
    \Phi(\omega^*,\bk)=\Phi^*(\omega,\bk).
\end{equation}
There is a clear physical reason: in the fast limit, the resonance condition
\begin{equation}
    \omega_\bk-\bk\cdot \bv=0
\end{equation}
cannot be satisfied for superluminal flavomons. Therefore, the integration contour in the dispersion relation need not be deformed, since it never encounters a pole in the first place---there are no neutrinos that can emit a flavomon given the kinematics of the reaction. Since the deformation of the contour is the only part of the integration that can endow $\Phi$ with an imaginary part for real $\omega_\bk$, we conclude that the dispersion relation is real-symmetric.

Therefore, an unstable or damped mode, meaning that $\mathrm{Im}(\omega_\bk)\neq 0$, cannot appear continuously as a function of $\bk$ from a stable branch with $\mathrm{Im}(\omega_\bk)=0$. A stable branch has
$\Phi(\omega_{R,\bk},\bk)=0$, where we denote by the suffix $R$ and $I$ the real and imaginary parts: if this condition is fulfilled, $\omega_{R,\bk}$ is a real-valued, and thus stable, flavor-wave mode. Since the dispersion relation itself is analytic, we may consider what happens at an infinitely close point $\bk+\delta \bk$ to the frequency $\omega_\bk+\delta \omega$, where the dispersion relation takes the form
\begin{equation}\label{eq:local_dispersion}
    \partial_\omega \Phi \delta \omega+\partial_\bk\Phi \cdot \delta \bk + \frac{1}{2}\partial_\omega^2 \Phi \delta \omega^2=0;
\end{equation}
the reason we keep the second derivative with respect to $\omega$ will become apparent in a moment. 

Since $\omega_\bk$ is purely real, and so is the function $\Phi(\omega,\bk)$, then for a small displacement $\delta \bk$ the mode itself remains stable with a frequency $\omega_{\bk+\delta\bk}=\omega_\bk+\delta\omega$ and
\begin{equation}
    \delta \omega=-\frac{\partial_\bk \Phi \cdot \delta \bk}{\partial_\omega \Phi}.
\end{equation}
This conclusions confirms that a stable superluminal branch remains superluminal, provided that $\partial_\omega \Phi\neq 0$. However, if, for some $\bk$, one point is crossed where this derivative vanishes, the branch must behave singularly. The condition therefore is the existence of a value of $\bk$ where we have simultaneously $\Phi(\omega,\bk)=0$ and $\partial_\omega \Phi(\omega,\bk)=0$. Since the first derivative vanishes, this condition implies the merging of two independent solutions of the dispersion relation, i.e.\ a degeneracy point.

An argument that was not made in Refs.~\cite{Yi:2019hrp,Fiorillo:2024uki}, and that we present here, is the counting of parameters, that allows us to determine where these degeneracies are expected to occur. Since there are four parameters $\bk$ and $\omega$, the two conditions that both the (real-valued) function $\Phi$ and its derivative vanish can be simultaneously satisfied on a two-dimensional surface in the space of wavevectors $\bk$. Therefore, degeneracies in the fast limit are expected to occur on two-dimensional surfaces. We note that the counting argument can be generalized; for example, in principle, there might be points where not only the first derivative, but also the second derivative $\partial_\omega^2 \Phi=0$. This condition corresponds to the degeneracy of three branches; the additional condition implies that this situation might only arise on one-dimensional lines in $\bk$ space, corresponding to the intersection of multiple degeneracy surfaces. We are not aware of any such case in the dispersion relation of fast flavor waves, but nevertheless it is possible in principle. We note here the formal similarity with the theory of phase transitions, where the equality of chemical potentials corresponding to different phases may lead to lines of phase transition between two phases, and triple points with transition between three phases.

In the vicinity of a degeneracy point, the local branch structure changes considerably; rather than having a locally linear dispersion $\delta\omega\propto \delta\bk$, we see from Eq.~\ref{eq:local_dispersion} that 
\begin{equation}
    \delta\omega=\pm\sqrt{-\frac{2 \partial_\bk \Phi \cdot \delta\bk}{\partial_\omega^2\Phi}}.
\end{equation}
Thus, $\delta\omega$ is real on one side of the surface, and purely imaginary on the other side of it. In practice, we have two complex conjugate modes on one side, and two real modes on the other side, becoming degenerate exactly on the surface. An unstable mode can therefore exist for superluminal velocities, together with its complex conjugate partner, appearing out of the merging of a pair of real modes on a surface of degeneracy.

For axially symmetric neutrino distributions, and for modes directed along the axis of symmetry, there is only a single parameter $k_z$. In other words, we look at the intersection of the degeneracy surface with the symmetry axis $\bk=k_z\hat{\bz}$. Therefore, the surface of degeneracy reduces along this axis only to a point. For this reason, the conventional axially symmetric examples studied in the literature (e.g.\ Refs.~\cite{Izaguirre:2016gsx,Capozzi:2017gqd,Airen:2018nvp,Yi:2019hrp,Fiorillo:2024uki,Fiorillo:2024pns,Fiorillo:2024dik,Fiorillo:2025ank}) retrieve these degeneracies always in the form of isolated points, which are visible e.g.\ in Fig.~5, case G3 in Ref.~\cite{Fiorillo:2024dik}. Here, the characteristic feature of the degeneracy point is a vertical tangent $d\omega/dk_z\to \infty$, which indeed is a direct prediction of the square-root behavior from our local analysis.

\subsection{General properties of degeneracy points}\label{sec:general_properties}

So far we have maintained the assumption that $\Phi(\omega_\bk,\bk)$ is purely real for $\mathrm{Im}(\omega_\bk)=0$. This assumption certainly fails when $|\omega_{R,\bk}|/|\bk|<1$, below the light cone as we usually say. Even worse, when neutrino masses are not neglected, there is not an individual light cone, since the resonance condition becomes
\begin{equation}\label{eq:resonance_condition}
    \omega-\bk\cdot \bv-w=0,
\end{equation}
so that even superluminal modes with $|\omega_R|>|\bk|$ may be resonant with some neutrinos for sufficiently large $w$. In practice, in realistic conditions $w$ is orders of magnitude smaller than the neutrino--neutrino interaction strength, as we have seen, but nevertheless at any given frequency there will be a small amount of neutrinos which can still satisfy the resonance condition. We aim to determine whether the degeneracy survives in these conditions, when the dispersion relation is not real-symmetric. 

We return then to Eq.~\ref{eq:dispersion_relation_formal}, describing the local behavior of a branch of flavor waves around a given solution $\bk$ and $\omega_\bk$. The only difference is that now even for real $\omega_\bk$, the function $\Phi(\omega_\bk,\bk)$ is complex. The general argument we have given remains true; a degeneracy point requires $\partial_\omega \Phi=0$ and $\Phi=0$ simultaneously. What changes, however, is the counting of parameters that determines whether these conditions can be satisfied.

In this case, the two complex equations are truly four equations, corresponding to the real and imaginary part of each of them. The parameters appearing in them are the three components of $\bk$, as well as the real and imaginary part of $\omega_\bk=\omega_{R,\bk}+i \omega_{I,\bk}$. For any given $\bk$, therefore, the additional conditions $\partial_\omega \Phi=0$ imposes two further constraints to be satisfied with three parameters. So below the light cone degeneracy can only occur on a line, rather than on a surface.
We come to the conclusion that a degeneracy surface in the massless neutrino approximation must be almost entirely lifted when neutrino masses are included; the degeneracy can survive at most on a line along this surface.

For the special case of axial symmetry, where usually only a single component $k_z$ is included, there is a single parameter, which cannot generically satisfy the two conditions $\partial_\omega\Phi=0$. Phrased another way, the degeneracy line which might occur in the three-dimensional space of wavevectors $\bk$, has vanishing probability of intersecting the symmetry axis---two lines in a three-dimensional space do not generically meet.
This is the direct explanation for why, in all previous studies~\cite{Izaguirre:2016gsx,Capozzi:2017gqd,Airen:2018nvp,Yi:2019hrp,Fiorillo:2024uki,Fiorillo:2024pns,Fiorillo:2024dik,Fiorillo:2025ank}, a degeneracy point below the light cone, where the dispersion relation is not real-symmetric, was never found.

Outside the light cone, a degeneracy point along the $k_z$ axis in the fast neutrino limit must be generically lifted when nonzero neutrino masses are included, barring fine-tuned conditions. This feature can be seen explicitly. In the fast limit, a degeneracy point corresponds to the formal condition
\begin{equation}
    \Phi_R(\omega,k_z)=0,\; \partial_\omega \Phi_R(\omega,k_z)=0.
\end{equation}
A nonzero vacuum frequency introduces a small imaginary part for the function $\Phi$, so that $\omega_{k_z}$ is no longer a solution. (There is also a slight shift of the real part of $\Phi$, which however does not impact qualitatively the degeneracy, and can be adjusted by a small change in $k_z$.) At a fixed value of $k_z$, we may look for the new solution as $\omega_{k_z}+\delta \omega$, and expand the dispersion relation as
\begin{equation}
    \frac{1}{2}\partial_\omega^2 \Phi_R(\omega_{k_z}, k_z)\delta \omega^2+ i\Phi_I(\omega_{k_z},k_z)=0,
\end{equation}
so that the single solution $\omega$ splits into two solutions with
\begin{equation}\label{eq:gap_opened}
    \delta \omega=\pm \sqrt{\frac{-2i \Phi_I(\omega_{k_z},k_z)}{\partial_\omega^2 \Phi_R(\omega_{k_z},k_z)}}.
\end{equation}
Therefore, the degeneracy point generically becomes gapped, i.e. separated in frequency, when nonzero neutrino masses are included. The entity of this frequency gap grows with the square root of the imaginary part of $\Phi(\omega,k_z)$ at the superluminal point, which in turn depends on the amount of DLN at frequencies $w$ large enough to satisfy the resonance condition in Eq.~\ref{eq:resonance_condition}. For example, the integrals $I_n$ develop an imaginary part
\begin{equation}
    \mathrm{Im}(I_n)=-\sqrt{2}G_F\pi\int_{-1}^{+1}dv v^n D_{v,\omega-k_z v},
\end{equation}
and therefore depend on the DLN at $w=\omega-k_z v\sim \mu\epsilon$. So the gap opening depends rather sensitively on the tail of the DLN distribution at large $w$, or on the neutrino and antineutrino distribution at small energies $E_\bp$.

\section{Termination singularities}\label{sec:landau_damping}

We now turn to the second kind of singularity in the dependence of $\omega_\bk$ on $\bk$, namely the abrupt termination of a solution. Such termination singularities appear when the solution approaches a singular point of the flavor dielectric function. They are connected at the mathematical level with the analytical continuation from the retarded prescription, and at the physical level with the nature of Landau damping as phase-space decoherence.

\subsection{Fast limit}\label{sec:abrupt_fast}

\begin{figure*}
    \includegraphics[width=\textwidth]{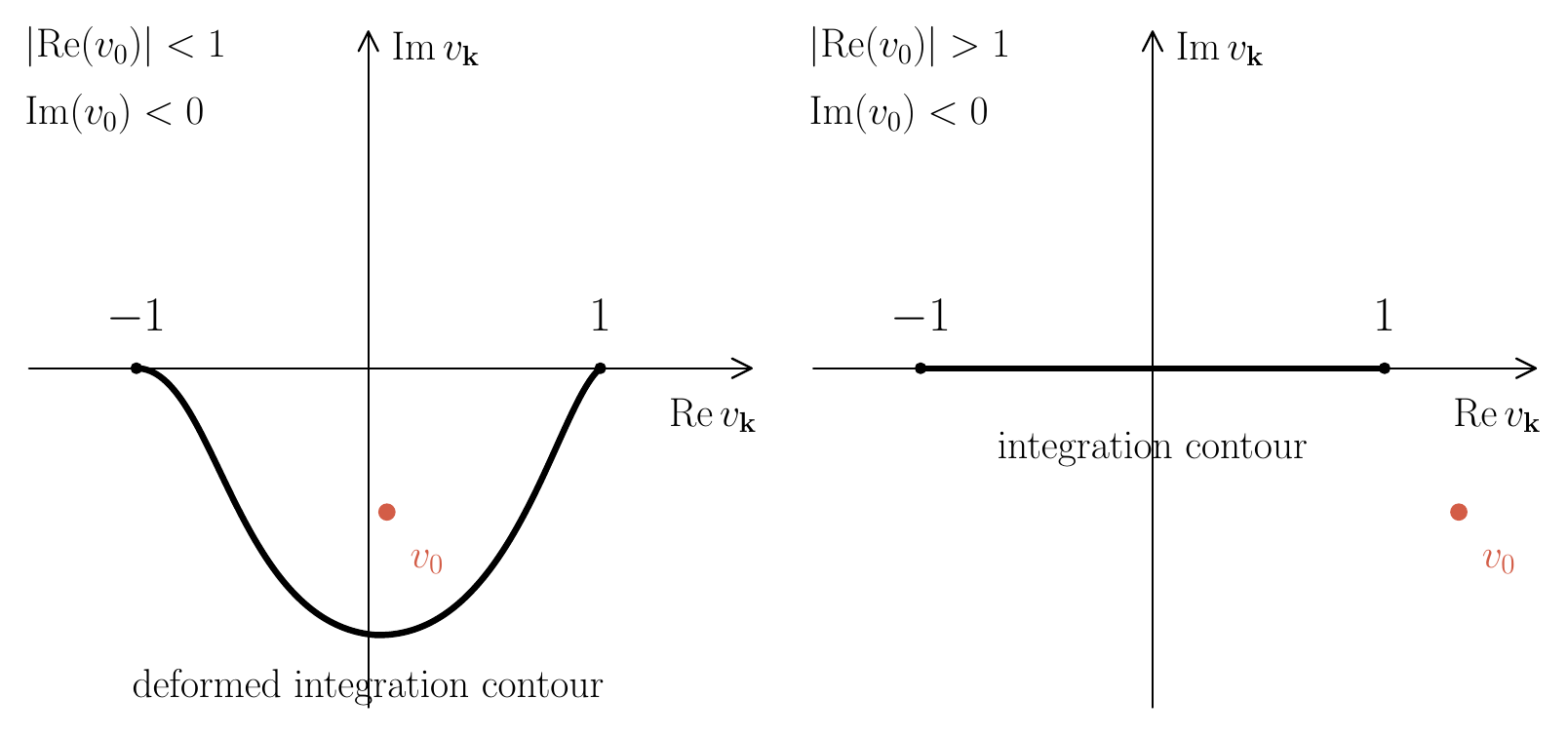}
    \caption{Deformation of the contour to perform the analytical continuation for the retarded dispersion relation, for subluminal (left) and superluminal (right) phase velocity $v_0=\omega/k$.}\label{fig:fast_deformation}
\end{figure*}

We begin our discussion with the fast limit of vanishing neutrino masses $w\to 0$. Once again, this case has been discussed already in Refs.~\cite{Fiorillo:2024bzm,Fiorillo:2024uki}, so we mostly summarize the results. 

In the fast limit, the analytical properties of the function $\Phi(\omega,\bk)$ is determined by the elementary integrals
\begin{equation}\label{eq:Imunu}
    I^{\mu\nu}(\omega, \bk)=\sum_\bp \frac{D_\bp v^\mu v^\nu}{\omega-\bk\cdot\bv+i0}.
\end{equation}
Since they all share the same analytical properties, it suffices to consider each one of them rather than their combination appearing in $\Phi(\omega,\bk)$.  We may perform the integration over the momentum of $\bp$ and its azimuthal angle around the axis of $\bk$, so that ultimately we only need to consider the one-dimensional integral in terms of the module $k=|\bk|$
\begin{equation}
    I^{\mu\nu}(\omega,k)=\int_{-1}^{+1} dv_\bk \frac{F^{\mu\nu}(v_\bk)}{\omega-k v_\bk+i0};
\end{equation}
notice that we are not assuming here axial symmetry, and we are merely changing the names of the variables to perform last the integration over the angle formed with $\bk$, $v_\bk=\cos\theta_\bk$. 

As discussed, the integral over $v_\bk$ is performed over a contour distorted to pass below the singularity $v_\bk=\omega/k$. For $\mathrm{Im}(\omega)>0$, no deformation is required, as the real axis already satisfies this property. For $\mathrm{Im}(\omega)<0$, the situation differs for subluminal ($|\omega|<k$) and superluminal ($|\omega|>k$) modes. 

For superluminal modes, even when the singularity moves in the lower half-plane, it still never intersects the contour of integration (right panel of Fig.~\ref{fig:fast_deformation}), since it lies outside of the range $|\mathrm{Re}(v_\bk)|<1$. Therefore, no deformation of the contour is required; this corresponds to the real-symmetric property that we have used in Sec.~\ref{sec:degeneracy_superluminal}. Instead, for subluminal modes, the contour should be deformed as to pass below the singularity, as in the left panel of Fig.~\ref{fig:fast_deformation}.

\begin{figure}
    \centering
    \includegraphics[width=0.8\textwidth]{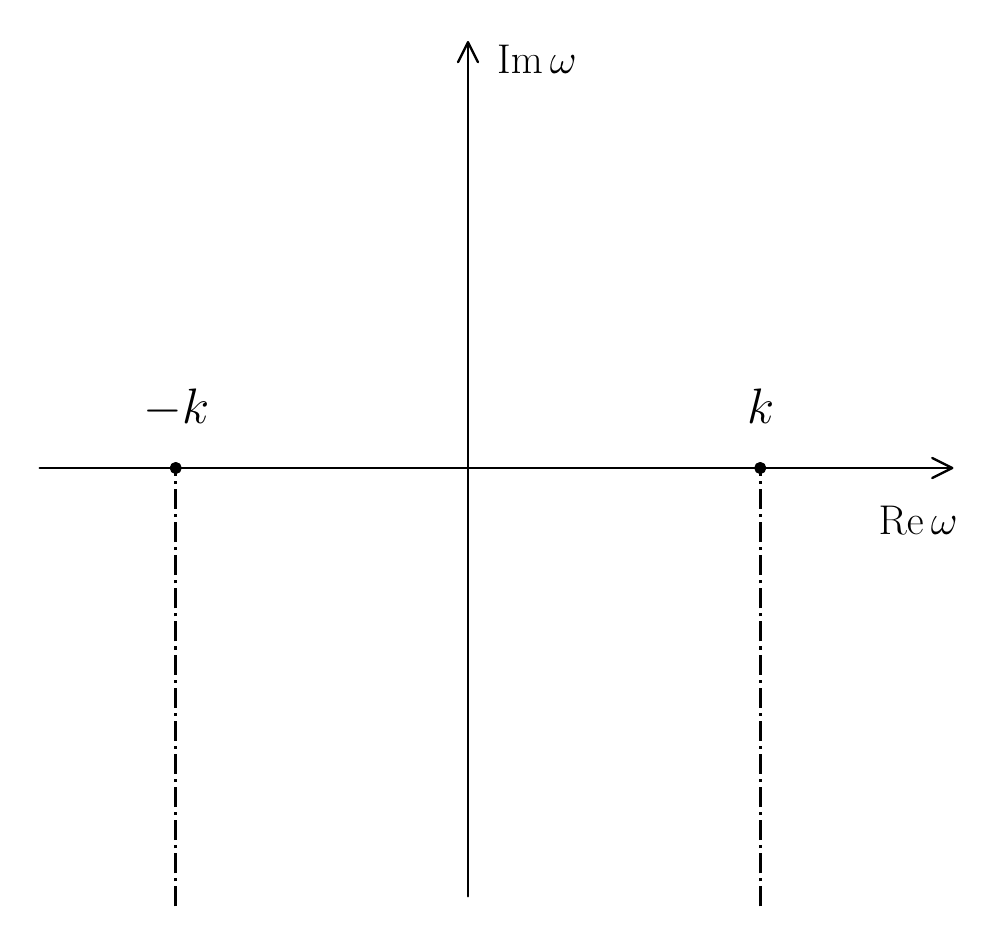}
    \caption{Branch cuts in the dependence of $\Phi(\omega,\bk)$ in the complex plane of $\omega$.}
    \label{fig:branch_cuts}
\end{figure}

The different integration prescription results directly in a singularity in the dependence on $\omega$. Since this discontinuity extends to all values of $\mathrm{Im}(\omega)<0$ corresponding to $\mathrm{Re}(\omega)=\pm k$, it takes the form of two branch cuts, as shown in Fig.~\ref{fig:branch_cuts}. These branch cuts effectively mark the light cone, where the phase velocity of flavor waves is the speed of light. 

For this reason a Landau-damped solution, existing between $-k<\mathrm{Re}(\omega)<k$, can terminate abruptly as it approaches the branch cut, and so can a superluminal damped solution with $|\mathrm{Re}(\omega)|>k$. This abrupt termination was ubiquitously found in the numerical solutions of the dispersion relation presented e.g.\ in Ref.~\cite{Fiorillo:2024dik}: for example, case G1 in Fig.~5 therein exhibits it rather clearly, with the two Landau-damped branches terminating as they touch the light cone. 

This phenomenon reflects the physical origin of Landau damping: the loss of coherence among neutrinos with neighboring directions. For a subluminal flavor wave, neutrinos whose velocity component along the propagation direction slightly exceeds the phase velocity move ahead in phase, while slightly slower neutrinos fall behind. Neutrinos near resonance remain nearly in phase with the wave over long times, allowing a sustained exchange of energy. For a superluminal wave, however, all neutrinos fall behind: none can satisfy the resonance condition, and Landau damping ceases.

\subsection{Nonzero neutrino masses}\label{sec:abrupt_slow}

As nonzero neutrino masses are included, the fate of Landau-damped modes becomes much less clear, because there is no individual light cone: even superluminal flavomons, as we have seen in Sec.~\ref{sec:general_properties}, can resonate with neutrinos with a sufficiently large $|w|$.

This issue has been noted recently in Ref.~\cite{Kost:2026jrk}, which proposes that the abrupt termination of solutions may be entirely removed when nonzero neutrino masses are included. We now show that this conclusion is true only for certain special cases of neutrino frequency distribution, but not in general.

\subsubsection{Abrupt termination from pinching singularities}\label{sec:pinching_singularities}

From our discussion on the fast limit we know that a solution $\omega_\bk$ can abruptly disappear as a function of $\bk$ only when it touches a singularity of the function $\Phi(\omega,\bk)$, or, what is the same, of the elementary integrals $I^{\mu\nu}$ defined in Eq.~\ref{eq:Imunu} entering that function. Let us therefore consider a generic integral of the form
\begin{equation}
    \mathcal{I}=\int_{-\infty}^{+\infty} dw \int_{-1}^{+1} dv_\bk \frac{F(v_\bk,w)}{\omega-kv_\bk -w+i0},
\end{equation}
where have integrated over the azimuthal angle of $\bv$ around the axis of $\bk$. The key challenge is to determine how to deform the contour of integration when $\mathrm{Im}(\omega)<0$, which is not trivial due to the integrations over multiple variables. 

As noted in Ref.~\cite{Kost:2026jrk}, there are multiple ways to perform this integral. One could integrate over $v_\bk$ first, as proposed in Ref.~\cite{Fiorillo:2025zio}, or one could integrate over $w$ first, or one could change variable to $w'=w+kv_\bk$ and integrate over $v_\bk$ first, as proposed in Ref.~\cite{Kost:2026jrk}. The specific procedure used cannot change the result of the integral, as shown explicitly in the Erratum to Ref.~\cite{Fiorillo:2025zio}, but the result may appear differently. For the purpose of our formal proof, it is much more convenient to use the integration procedure proposed in Ref.~\cite{Kost:2026jrk}.

\begin{figure*}
    \includegraphics[width=\textwidth]{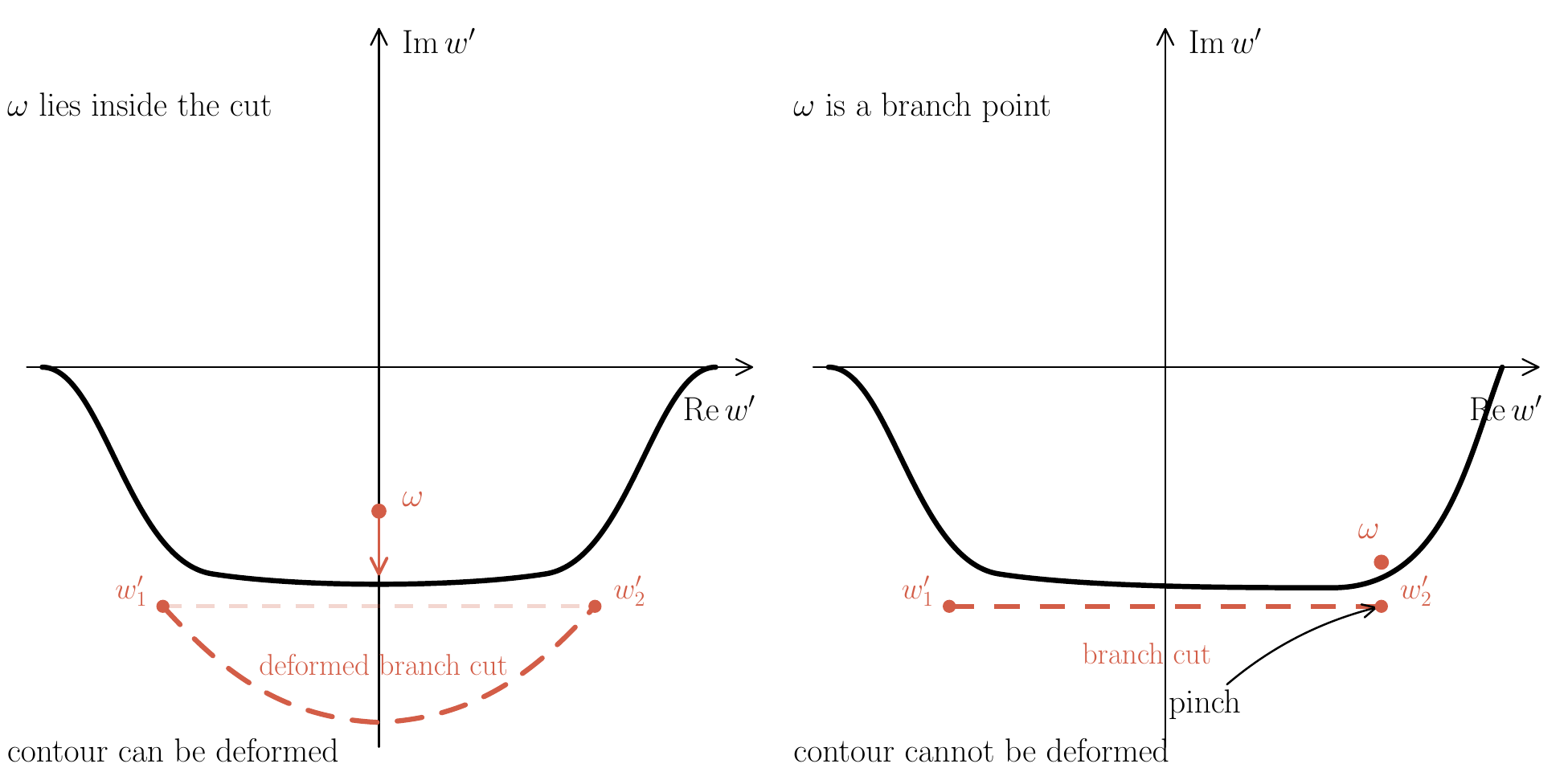}
    \caption{ Deformation of the contour in $w'=w+kv_\bk$ when the function $G(w')$ has a branch-cut singularity in the lower half-plane with branch points $w'_1$ and $w'_2$. In the left panel, even when $\mathrm{Im}(\omega)$ becomes lower than the branch cut, the contour can still be deformed by pushing the branch cut downward, so $\mathcal{I}$ has no singularity for $\mathrm{Re}(w'_1)<\mathrm{Re}(\omega)<\mathrm{Re}(w'_2$). In the right panel, the singularity at $\mathrm{Re}(\omega)=\mathrm{Re}(w'_1)$ or $\mathrm{Re}(\omega)=\mathrm{Re}(w'_2)$ cannot be removed, since the branch cut is pinned to the endpoints. The contour is thus pinched between the resonant pole $w'=\omega$ and the branch point $w'_2$.}\label{fig:endpoint_pinch}
\end{figure*}

With this change of variable, the integral
\begin{equation}
    \mathcal{I}=\int_{-\infty}^{+\infty}\frac{dw'}{\omega-w'+i0}\int_{-1}^{+1}dv_\bk F(v_\bk,w'-kv_\bk)=\int_{-\infty}^{+\infty}dw' \frac{G(w')}{\omega-w'+i0}.
\end{equation}
Since there is no discontinuous behavior in the range of integration of $w'$, it might appear now that there are no singularities; for any value of $\omega$, the contour must be deformed to pass below the singularity $w'=\omega$ if $\mathrm{Im}(\omega)<0$. However, this deformation can only be performed if the function $G(w')$ has no singularities in the lower half-plane. 

If $G(w')$ has a singularity in the lower half-plane, say, a branch cut as in Fig.~\ref{fig:endpoint_pinch}, then the deformation faces an obstacle. If the pole $w'=\omega$ is within the branch cut, as in the left panel, the deformation can still be carried out, since the branch cut can be arbitrarily deformed in the lower half-plane as in figure. However, if the pole touches one of the two branch points, as in the right panel of the figure, no deformation of the branch cut can allow the integral to be performed. We thus have a pinch singularity; the integration contour is pinched between the pole $w'=\omega$ and a branch point of the integrand function. 

We reach the conclusion that the singularities of the dispersion relation are determined by the singularities of $G(w')$ in the lower half-plane. $G(w')$ itself is defined as
\begin{equation}
    G(w')=\int_{-1}^{+1}dv_\bk F(v_\bk,w'-kv_\bk),
\end{equation}
so its singularities are determined in turn from those of $F(v_\bk,w'-kv_\bk)$ in the following way. If $F(v_\bk,w)$ has a pole or a branch point at $w_s(v_\bk)$, the function $G(w')$ develops as a result two branch points respectively at $w'_\pm=w_s(\pm1)\pm k$. We can see this rather easily by considering the local structure of a generic branch point $F(v,w)\propto (w-w_s(v_\bk))^\alpha$. For $\alpha=-1$, this singularity turns into a pole, so our argument below is quite generic. The integral
\begin{equation}
    \int_{-1}^{+1}dv_\bk (w'-kv_\bk-w_s(v_\bk))^\alpha
\end{equation}
receives contribution from the endpoints in proportion to $(w'\mp k-w_s(\pm 1))^{\alpha+1}$, so $w'_\pm$ become branch points. (For $\alpha=-2, -3,...$ the integral produces poles rather than branch cuts, although this is a somewhat specialized case.) Another kind of singularity can arise if the derivative $k+w_s'(v^*)$ vanishes within the range $-1<v^*<1$, in which case $w^{'*}=k v^*+w_s(v^*)$ becomes another singularity of $G(w')$; as a simple example the function
\begin{equation}
    \int_{-1}^{+1}dv (w'-v^2)^\alpha
\end{equation}
is singular for $w'=0$. 

We thus reach the following, general result: if the neutrino DLN distribution $F(v_\bk,w)$ has singularities at $w=w_s(v_\bk)$ of any kind, either branch points or poles, the dispersion relation will develop two singularities of the branch-cut type at
\begin{equation}
    \mathrm{Re}(\omega)=\pm k + \mathrm{Re}[w_s(\pm1)],\; \mathrm{Im}(\omega)<\mathrm{Im}[w_s(\pm 1)].
\end{equation}
These two branch cuts are the direct heritage of the branch cuts in the fast case, where they were located at $\mathrm{Re}(\omega)=\pm k$, $\mathrm{Im}(\omega)<0$. Another special singularity, less interesting in practice, appears for $\omega=k v^*+w_s(v^*)$, as we have seen.

On the other hand, Ref.~\cite{Kost:2026jrk} presents explicit examples in which no singularity appears, and the functions $\omega_\bk$ are completely continuous. The reason, in light of the previous argument, is that their examples use Gaussian frequency distributions for $F(v_\bk,w)$. A Gaussian distribution is an entire function, which was the motivation for their choice in Ref.~\cite{Kost:2026jrk}, namely it possesses no singularity. In turn, the function $\Phi(\omega,\bk)$ also has no singularity. However, this property hinges on the specific choice of function. As we will see, even slight deformations of the distribution that can look quite similar to the Gaussian close to the peak can have entirely different behaviors and lead to singularities.

\subsubsection{Lorentzian distributions as an explicit example}

The above discussion is quite technical, so it is more instructive to see the appearance of singularities with a specific example. A particularly convenient example is a Lorentzian distribution of frequencies, so we focus on the analytical properties of integrals of the form
\begin{equation}
    \mathcal{I}=\int_{-\infty}^{+\infty}dw \int_{-1}^{+1}dv_\bk \frac{\Gamma/\pi}{(w-w_0)^2+\Gamma^2}\frac{f(v_\bk)}{\omega-kv_\bk-w+i0}.
\end{equation}
The reason that this choice is so instructive is that the integral over $w$ can be performed analytically. The integrand function has three poles, at $w=\omega-kv$ and $w=w_0\pm i\Gamma$. The integration contour always lies below the first pole. Similarly, the second pole $w=w_0+i\Gamma$ also lies above the integration contour. Therefore, if we close the integration contour in the lower half-plane, the integration catches only the contribution from the third pole $w=w_0-i\Gamma$, and
\begin{equation}\label{eq:dispersion_lorentzian_beam}
    \mathcal{I}=\int_{-1}^{+1}dv_\bk \frac{f(v_\bk)}{\omega-kv_\bk-w_0+i\Gamma}.
\end{equation}
Thus we discover a remarkable result, namely that a Lorentzian distribution of vacuum frequencies, from the perspective of the dispersion relation, behaves exactly as a monochromatic neutrino beam with frequency $w_0$ and a collision rate $\Gamma$. 
This result descends entirely from the retarded prescription of integration, as is clear from the fact that Eq.~\ref{eq:dispersion_lorentzian_beam} is complex even for real $\omega$.

From Eq.~\ref{eq:dispersion_lorentzian_beam} we now see explicitly the singularity structure; a solution can disappear whenever $\mathrm{Im}(\omega)<-\Gamma$ and $\mathrm{Re}(\mathrm{\omega})=w_0\pm k$. So the role of the vacuum frequency distribution is merely to displace the singularity of the light cone, but not to destroy it. This conclusion might appear at first sight quite surprising, in light of our previous statement that a Gaussian distribution does not have any singularity. On the real axis, a Gaussian and Lorentzian distribution appear rather similar close to their peaks. It is instructive to prepare an explicit example, to compare the dispersion relation solved for these two cases, so that one can observe explicitly their remarkably different behavior.

In order to simplify the example to a form that can be followed easily, we do not use the full dispersion relation $\Phi(\omega,\bk)$ from Eq.~\ref{eq:dispersion_relation_formal}, which involves multiple modes through the determinant of the flavor dielectric function. Instead, we construct directly the local structure of the dispersion relation close to a point where a Landau-damped mode disappears. This local structure should allow, in the fast limit, for a solution with $\mathrm{Im}(\omega)<0$ and $|\mathrm{Re}(\omega)|<k$. For positive $0<\mathrm{Re}(\omega)<k$, the solution of the dispersion relation in the fast limit should therefore have a phase $z=\omega-k=|z| e^{i\phi}$ with $\phi\to 3\pi/2$. In the fast limit, we thus use a dispersion relation
\begin{equation}
    \phi(\omega,k)=\int_{-1}^{+1}dv_\bk \frac{D}{\omega-kv_\bk+i0}-f=0,
\end{equation}
where for simplicity we assume an isotropic distribution $D_{v_\bk}=D$. The constant $f$ should be chosen such that at $k=k^*$ there is a Landau-damped solution with $\mathrm{Re}(\omega)\to k^*$ from the left side, and $\mathrm{Im}(\omega)=-\gamma^*$. Performing the integral, and following the retarded prescription such that
\begin{equation}
    \log(-i\gamma)=\log(\gamma)+\frac{3i\pi}{2},
\end{equation}
we see that
\begin{equation}
    f=\frac{D}{k^*}\left[\log(2k^*-i\gamma)-\log(\gamma)-\frac{3i\pi}{2}\right]=\frac{D}{k^*}\left[\log\left(\sqrt{1+\frac{4 k^{*2}}{\gamma^2}}\right)-i \arctan\left(\frac{\gamma}{2k^*}\right)-\frac{3i\pi}{2}\right].
\end{equation}

We can now solve and compare three different dispersion relations:
\begin{equation}
    \phi_{\rm fast}(\omega,k)=\int_{-1}^{+1}dv_\bk \frac{D}{\omega-kv_\bk+i0}-f=0
\end{equation}
which does not include any effect from the vacuum frequency;
\begin{equation}
    \phi_{\rm Lor}(\omega,k)=\int_{-\infty}^{+\infty}dw \frac{\Gamma/\pi}{w^2+\Gamma^2}\int_{-1}^{+1}dv_\bk \frac{D}{\omega-kv_\bk-w+i0}-f=0,
\end{equation}
which includes a Lorentzian frequency distribution; and
\begin{equation}
    \phi_{\rm Gauss}(\omega,k)=\int_{-\infty}^{+\infty}dw\frac{e^{-\frac{w^2}{2\Gamma^2}}}{\sqrt{2\pi}\Gamma}\int_{-1}^{+1}dv_\bk\frac{D}{\omega-kv_\bk-w+i0}-f=0
\end{equation}
which includes a Gaussian frequency distribution. 

For the fast and the Lorentzian case, the integral can be performed explicitly; in the fast case, the integral gives a logarithm, from which we obtain as a solution
\begin{equation}
    \omega_{k,\rm fast}=k\coth\left(\frac{fk}{2D}\right).
\end{equation}
This expression however holds only for $k<k^*$, where the phase of $\omega-k$ is less than $3\pi/2$; when the phase of $\omega-k$ crosses $3\pi/2$, the solution is abruptly interrupted and does not satisfy $\phi_{\rm fast}=0$ because of the specific branch prescription of the logarithm, precisely as anticipated.

For the Lorentzian case, as we have seen, the integration over $w$ merely endows the monochromatic zero-frequency beam with a complex $w\to -i \Gamma$, so the solution is merely shifted
\begin{equation}
    \omega_{k,\rm Lor}=-i\Gamma+k\coth \left(\frac{fk}{2D}\right).
\end{equation}
For the Gaussian case, there is no simple analytical result; the integral over $w$ can be performed explicitly to give
\begin{equation}
    \phi_{\rm Gauss}(\omega,k)=-\int_{-1}^{+1}dv_\bk\frac{D}{\sqrt{2}\Gamma}Z\left(\frac{\omega-kv_\bk}{\sqrt{2}\Gamma}\right)-f=0,
\end{equation}
where
\begin{equation}
    Z(x)=i\sqrt{\pi}e^{-x^2}\mathrm{erfc}(-ix),
\end{equation}
and erfc$(z)$ is the complementary error function. The integral over $v_\bk$ needs to be performed numerically.

\begin{figure*}
    \includegraphics[width=\textwidth]{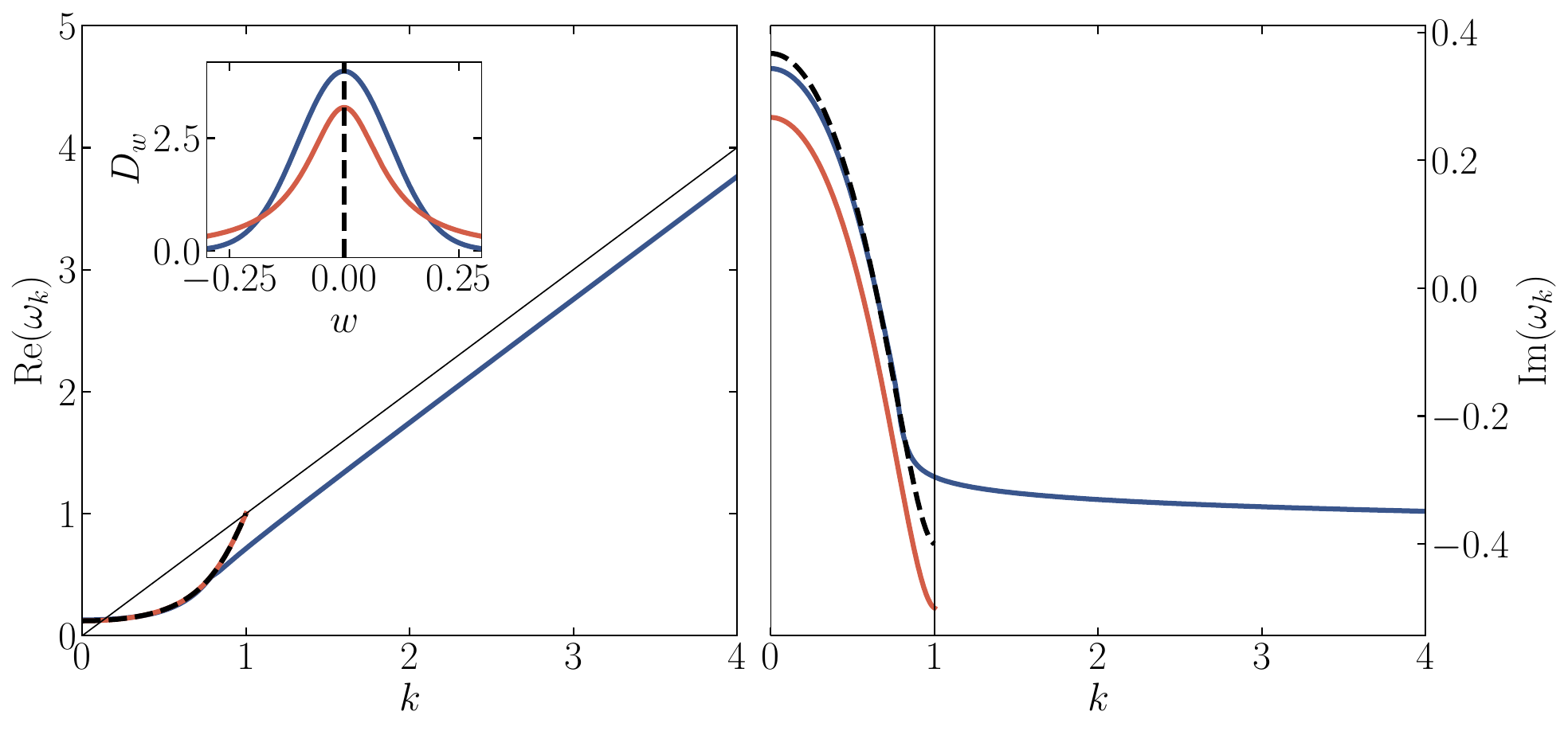}
    \caption{ Comparison between the solution of the dispersion relation assuming a monochromatic frequency distribution at $w=0$ (fast limit, black dashed), a Gaussian distribution with semidispersion $\Gamma=0.1$ (blue), and a Lorentzian distribution with width $\Gamma=0.1$ (red). The distributions are shown in the inset. We show the real (left) and imaginary (right) part of $\omega$ as a function of $k$. In the left panel, the red and black lines are superposed. The numerical values of the parameters adopted are $D=1$, $k^*=1$, $\gamma=0.4$. The light cone is shown as a thin black line.}\label{fig:gauss_lorentz}
\end{figure*}

We show an explicit numerical comparison in Fig.~\ref{fig:gauss_lorentz}: the numerical values adopted for the parameters are collected in the figure caption. As we expected from the formal discussion in Sec.~\ref{sec:pinching_singularities}, and from the results of Ref.~\cite{Kost:2026jrk}, the Gaussian distribution possesses a branch $\omega_k$ which is continuous for all $k$. For small $k$, its real part coincides with the solution of the dispersion relation for the fast limit and the Lorentzian distribution. However, the latter are abruptly terminated when they touch the light cone with a negative $\mathrm{Im}(\omega_k)$, in agreement with our discussion on pinching singularities.

Several features of this example serve to show explicitly the general properties we have highlighted before. For example, all three solutions pass through the light cone at small $k$ without any discontinuous behavior: the reason is that their $\mathrm{Im}(\omega_k)$ is in all cases positive, so there cannot be any singularity there---they do not meet the branch cut at $\mathrm{Im}(\omega)<\mathrm{Im}(w_{\rm pole})$. As another interesting feature, the Lorentzian result can be obtained from the fast limit simply by subtracting $\omega_k \to \omega_k-i\Gamma$. 

\subsection{Summary}

Given the somewhat technical nature of this section, it is useful to summarize its main content. We have found that termination singularities in the dispersion relation do not appear only in the fast limit of vanishing neutrino masses. Rather, they depend on the singularities of the neutrino frequency distribution. A Gaussian frequency distribution, as those considered in Ref.~\cite{Kost:2026jrk}, possesses no singularities, and therefore exhibits no discontinuous behavior in its flavor-wave dispersion. However, more generally one expects such discontinuities.

We have provided a general rule to detect singularities in the dispersion relation: if the angular and frequency distribution is singular at $w_s(v)$, then the dispersion relation will possess two branch cuts at
\begin{equation}
    \mathrm{Re}(\omega)=\pm k + \mathrm{Re}[w_s(\pm 1)],\; \mathrm{Im}(\omega)<\mathrm{Im}[w_s(\pm 1)].
\end{equation}
If there are coinciding singularities at the same value of $v^*$, i.e.\ simultaneous vanishing of $\omega=k v^* + w_s(v^*)$ and $k + dw_s(v^*)/dv^*=0$, an additional type of singularity appears, which we have not studied in more depth given its special nature.

In addition to the general rule, we have considered a specific example, to highlight how differently the Landau-damped modes behave even for frequency distributions that look relatively similar close to their peak. A Gaussian and Lorentzian distribution with a similar peak structure exhibit a completely different flavor-wave dispersion; the Gaussian distribution shows a Landau-damped mode continuing for arbitrary $k$, while a Lorentzian distribution shows the same Landau-damped mode disappearing abruptly as it touches the branch cut from its complex pole.

Finally, we have shown that Lorentzian frequency distributions are equivalent to monochromatic frequency distributions with a collision rate. This feature opens up a significant practical simplification in the solution of the dispersion relation, since it removes the need for frequency integration---such integrations, especially in view of the specific retarded prescription that needs to be followed, can be numerically time-consuming. While the true neutrino frequency distributions are certainly not Lorentzians in realistic scenarios, they can be represented as combinations of Lorentzians in the bulk region.

More generically, our results serve as a warning against overinterpreting the outcome of the dispersion relation at large negative $\mathrm{Im}(\omega_k)$; this outcome depends rather sensitively on the extension of the frequency and angular distribution to the complex plane, which can be completely different even for distributions which look remarkably close on the real axis.

\section{Explicit examples}\label{sec:examples}

We finally turn to the flavor-wave dispersion relation $\Phi(\omega,\bk)=0$, and show how the features we have identified appear in its solution. We consider two cases: a purely slow-unstable case, and a fast-unstable case in the presence of nonzero neutrino mass splittings.

\subsection{Slow instability}

\begin{figure*}
    \includegraphics[width=\textwidth]{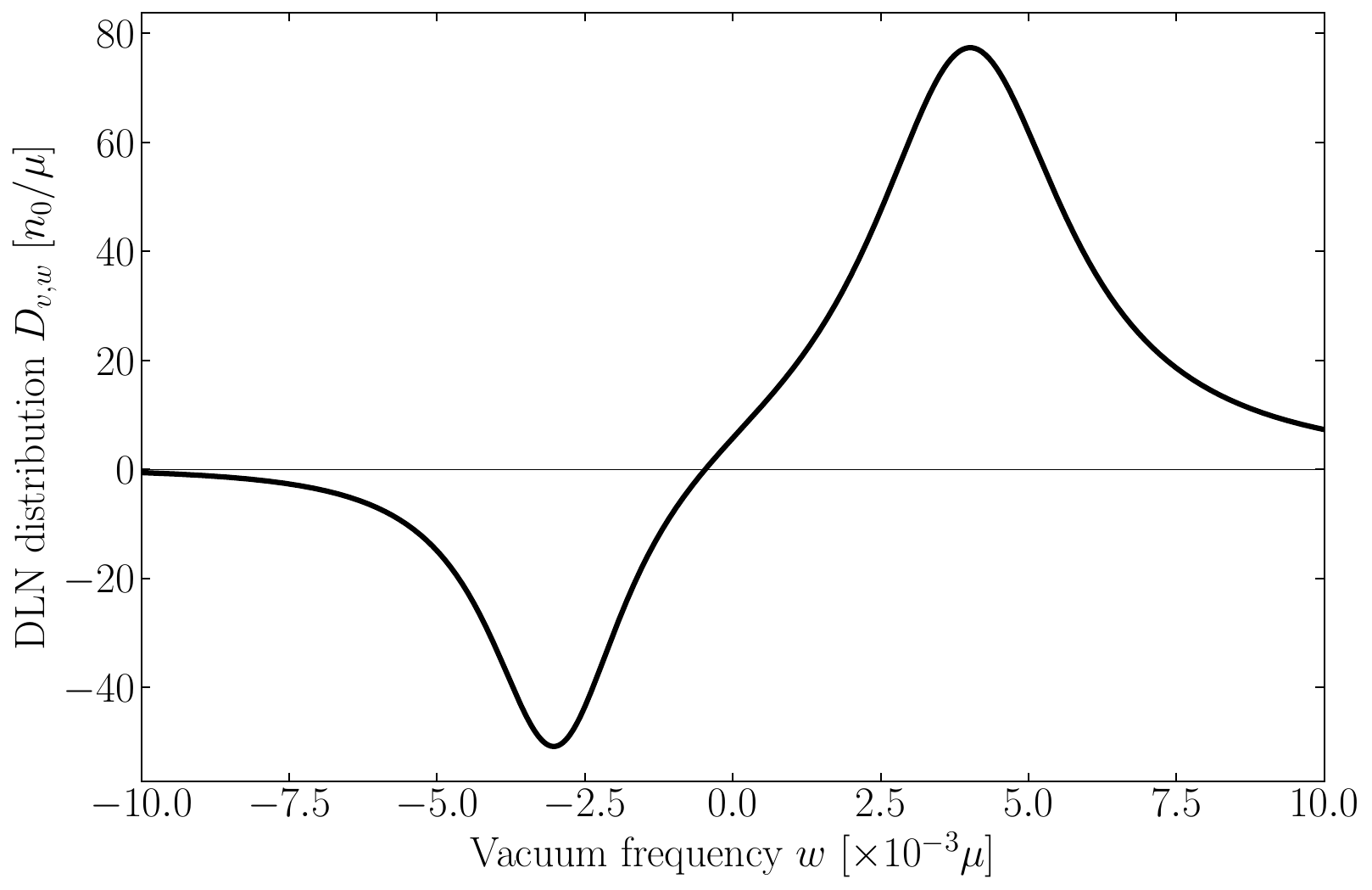}
    \caption{Benchmark DLN distribution used to determine the flavomon dispersion relation. The numerical values of the parameters adopted are $w_\nu=4\times 10^{-3}\mu$, $w_{\overline{\nu}}=-3\times 10^{-3}\mu$, $\Gamma_\nu=2\times10^{-3}\mu$, $\Gamma_{\overline{\nu}}=1.4\times 10^{-3}\mu$, $n_\nu=n_0$, $n_{\overline{\nu}}=0.5n_0$.}\label{fig:distribution}
\end{figure*}

For the case of a pure slow instability, the features we aim to show appear already for the simplest case of an isotropic neutrino distribution, which we therefore take as our example. In practice, we choose a frequency distribution with two Lorentzians with opposite signs
\begin{equation}
    D_{v,w}=\frac{1}{2\pi}\left[\frac{\Gamma_\nu n_\nu}{(w-w_\nu)^2+\Gamma_\nu^2}-\frac{\Gamma_{\overline{\nu}}n_{\overline{\nu}}}{(w-w_{\overline{\nu}})^2+\Gamma_{\overline{\nu}}^2}\right].
\end{equation}

Note the units of the quantities: $w_E$ is an energy, and so are the widths $\Gamma_\nu$ and $\Gamma_{\overline{\nu}}$, whereas $n_\nu$ and $n_{\overline{\nu}}$ are number densities. It is convenient to express them in units of a  reference number density $n_0$ such that $\mu=\sqrt{2}G_F n_0$, and to express all energies in units of $\mu$. We assume this convention in the following. The benchmark DLN distribution used for this study is shown in Fig.~\ref{fig:distribution}. Schematically, it represents a predominantly positive DLN at positive frequencies $w>0$, corresponding to a dominant $\nu_e$ population, and a predominantly negative DLN at negative frequencies $w<0$, corresponding to a dominant $\overline{\nu}_e$ population. The integrated DLN is positive, so that the antineutrinos $\overline{\nu}_e$ play the role of the flipped species~\cite{Fiorillo:2025npi}, i.e.\ they tend to emit rather than absorb flavomons.

Since the angular distribution is isotropic, the dispersion relation can always be taken to be that of an axially symmetric mode---any axis is an axis of azimuthal symmetry, or, in other words, the frequency $\omega_\bk$ depends only on the module of $\bk$. We consider only the dispersion relation for longitudinally polarized modes in Eq.~\ref{eq:dispersion_relation_longitudinal}. With our convention on the units, the integrals $I_n$ simply read
\begin{equation}
    I_n=\int_{-1}^{+1}dv\int_{-\infty}^{+\infty}dw \frac{D_{v,w}v^n}{\omega-k v - w+i0};
\end{equation}
the integral over $w$ can be performed explicitly to give
\begin{equation}
    I_n=\int_{-1}^{+1}\frac{dv}{2} v^n\left[\frac{n_\nu}{\omega-kv-w_\nu+i\Gamma_\nu}-\frac{n_{\overline{\nu}}}{\omega-kv-w_{\overline{\nu}}+i\Gamma_{\overline{\nu}}}\right].
\end{equation}

As a matter of fact, even the integrals over $v$ are exact, provided one adopts the retarded prescription for the integration. The result to use in this case for the family of integrals
\begin{equation}
    J_n=\int_{-1}^{+1}dv\frac{v^n}{\omega-kv-w+i\Gamma}
\end{equation}
is the recursion relation
\begin{equation}
    J_{n+1}=\frac{\omega-w+i\Gamma}{k}J_n-\frac{(1-(-1)^{n+1})}{k(n+1)}
\end{equation}
and
\begin{equation}
    J_0=\frac{1}{k}\left[\log(\omega+k-w+i\Gamma)-\log(\omega-k-w+i\Gamma)\right].
\end{equation}
Here the logarithm is defined with the branch prescription discussed e.g.\ in Refs.~\cite{Fiorillo:2024pns,Fiorillo:2025ank,Fiorillo:2025zio}, namely that 
\begin{equation}
    \log(z)=\log(|z|)+i\phi
\end{equation}
with $-\pi/2<\phi<3\pi/2$.
\begin{figure*}
    \includegraphics[width=\textwidth]{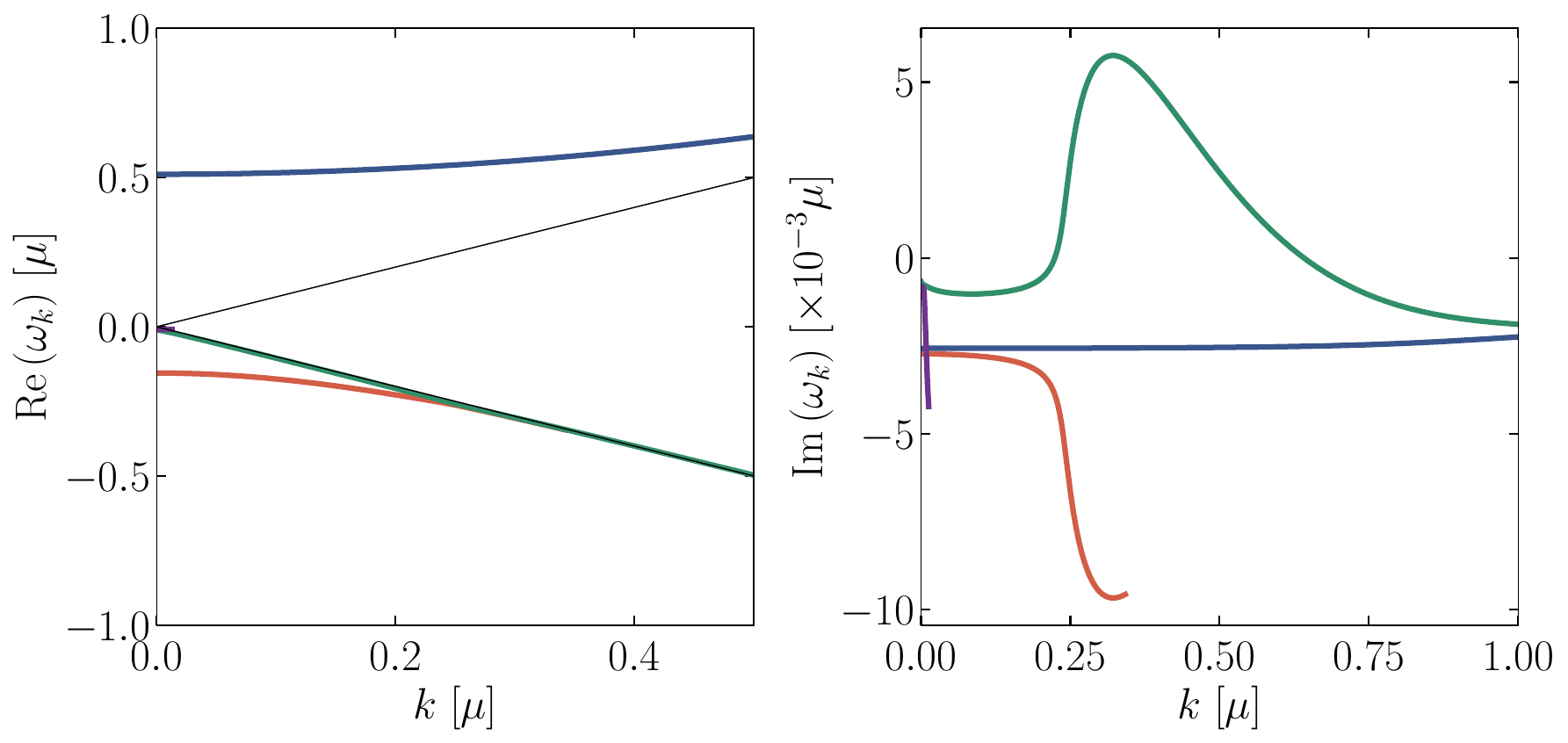}
    \caption{Real (left) and imaginary (right) part of the eigenfrequency as a function of wavevector $k=|\bk|$ for the benchmark distribution shown in Fig.~\ref{fig:distribution}. The light cone is shown in black; each of the four branches is shown in color. The red and purple branches abruptly terminate when they touch a singular branch cut of the dispersion relation.}\label{fig:dispersion_relation_slow}
\end{figure*}

With these results, the dispersion relation is entirely analytical; its solution of course should still be performed numerically. The results are shown in Fig.~\ref{fig:dispersion_relation_slow}.

We find overall four branches of solutions to the dispersion relation. The blue and red branch are the gapped branches~\cite{Fiorillo:2025zio}: they exist even in the fast limit $w\to 0$ and therefore always have $\omega_k(k=0)\sim \mu\epsilon$. The green and purple solution are gapless, in the sense that their frequency at small $k$ is of the order of $w$; they disappear when $w\to 0$. The green and red solution develop a nonzero growth rate, positive and negative respectively, as they approach the light cone. This conclusion matches all previous results on the subject~\cite{Fiorillo:2024pns,Fiorillo:2025ank,Fiorillo:2025zio,Fiorillo:2025kko}, and most importantly it matches the qualitative understanding~\cite{Fiorillo:2026gzy}. In normal ordering, the resonance condition $\omega-kv-w=0$ for flavomons with negative frequency $\omega<0$ can only be satisfied by neutrinos with negative frequency $w<0$, which in our case are the flipped antineutrinos, emitting, rather than absorbing, flavomons. Thus, a branch approaching the negative light cone exhibits an instability. Vice versa, the blue branch approaching the positive light cone resonates with neutrinos with a positive $w>0$, which in our benchmark distribution are neutrinos with a positive DLN, absorbing, rather than emitting, flavomons, so the modes do not turn unstable. The two striking features of the solution we find, in relation to our discussion of singularities, are the following. 

\begin{figure*}
    \includegraphics[width=\textwidth]{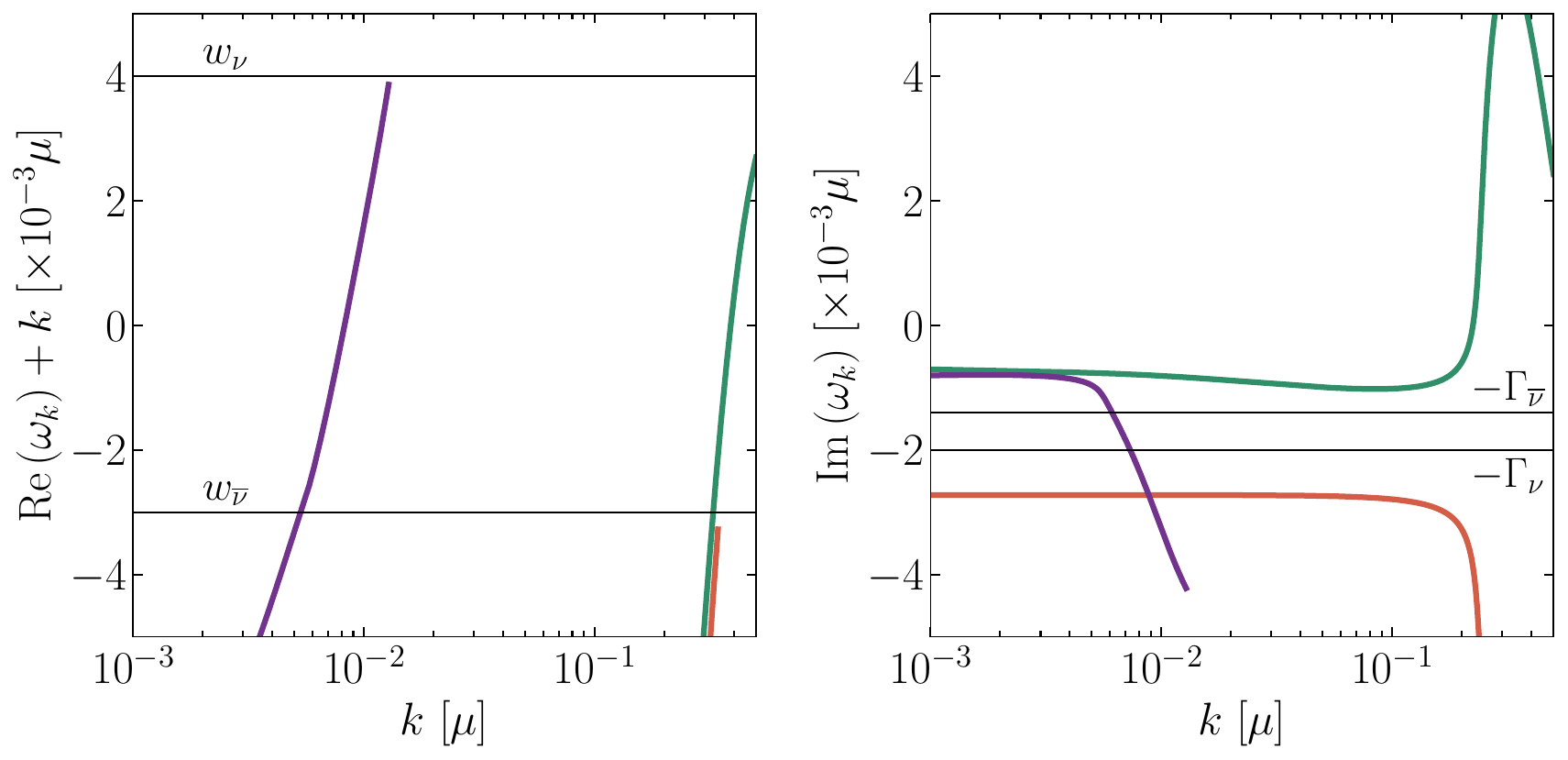}
    \caption{Zoomed-in version of Fig.~\ref{fig:dispersion_relation_slow}. For the real part, we show the distance from the light cone $\mathrm{Re}(\omega_k)+k$ rather than $\mathrm{Re}(\omega_k)$ itself. The horizontal lines are the real and imaginary part of the branch points predicted by the rules in Sec.~\ref{sec:abrupt_slow}; for a Lorentzian distribution, these are simply the centers and widths of the distributions.}\label{fig:dispersion_relation_slow_zoom}
\end{figure*}

First, no modes ever become degenerate. The red and green branch show a clear separation at all $k$. In order for us to show this feature rather clearly, it was essential that the integral over frequency could be performed analytically; the gap opening, as we have seen, depends on the DLN distribution at frequencies $w\sim \mu \epsilon$, in the far tail of the frequency distribution. In terms of neutrino energies, the relevant range is $E_\bp\sim \delta m^2/\mu\epsilon$, so in order to retrieve the gap opening a numerical integration should sample properly the energy distribution at very small energies of the order of tens of eV. This conclusion is the likely explanation for why such a gap was not observed in the numerical integration of Ref.~\cite{Fiorillo:2025zio}. Another interesting feature is that, for the gap opening to show up, we had to choose $\Gamma_\nu$ and $\Gamma_{\overline{\nu}}$ different. If they had been chosen equal to a common $\Gamma$, the frequency might simply have been redefined to $\omega_k \to \omega_k+i\Gamma$, so the dispersion relation reduces to that of two monochromatic beams with no width. For this case, the dispersion relation is real outside of the light cone, and therefore there is no gap opening according to the discussion in Sec.~\ref{sec:degeneracies}, and as also found in the examples of Ref.~\cite{Fiorillo:2024pns,Fiorillo:2025ank}. This feature is of course a mere accident of the choice of a common width, and serves as a reminder of how easy it is to choose specialized conditions exhibiting a mode structure not representative of the generic one.

The second feature that our benchmark examples exhibit rather clearly is the abrupt disapperance of modes: the purple branch disappears at very small $k$, while the red branch survives up to $k\sim 0.3\mu$ before disappearing. Thus, as we had predicted in Sec.~\ref{sec:abrupt_slow}, the abrupt disappearance of damped modes survives the presence of a nonzero vacuum frequency; the broadening of the light cone does not regularize it. Of course, to really put to the test the prediction in Sec.~\ref{sec:abrupt_slow}, we also need to check that the solutions abruptly disappear when they touch the branch cuts deduced by the rules in Sec.~\ref{sec:abrupt_slow}.

We do so in Fig.~\ref{fig:dispersion_relation_slow_zoom}, where we zoom in on the regions where the modes terminate abruptly. By showing $\mathrm{Re}(\omega_k)+k$, we see clearly that the modes disappear not at the light cone, which would be the position of the singularity in the fast limit, but rather at the branch cuts $\mathrm{Re}(\omega_k)+k=w_\nu$ (for the purple band) and $\mathrm{Re}(\omega_k)+k=w_{\overline{\nu}}$ (for the red band). We emphasize that this finding is purely numerical---the solution of the dispersion relation terminates abruptly at that position, with the root-finder being unable to find any solution for larger values of $k$---and matches entirely the theoretical prediction, while being independent of it.

Incidentally, the purple curve also exhibits an interesting effect, in that it crosses the line $\mathrm{Re}(\omega_k)+k=w_{\overline{\nu}}$ without disappearing. The reason can be gathered by looking at its imaginary part; at the value of $k$ where this crossing happens, its imaginary part $\mathrm{Im}(\omega_k)>-\Gamma_{\overline{\nu}}$. Therefore, in the complex plane, the purple band is not touching the branch cut, which, from the general rules we have deduced, lies at $\mathrm{Re}(\omega_k)+k=w_{\overline{\nu}}$ but also $\mathrm{Im}(\omega_k)<-\Gamma_{\overline{\nu}}$.

In summary, our numerical solution of the dispersion relation matches all of the general predictions based on the theorems in Secs.~\ref{sec:degeneracies} and~\ref{sec:landau_damping}, not just qualitatively but quantitatively.

\subsection{Fast instability}

\begin{figure*}
    \includegraphics[width=\textwidth]{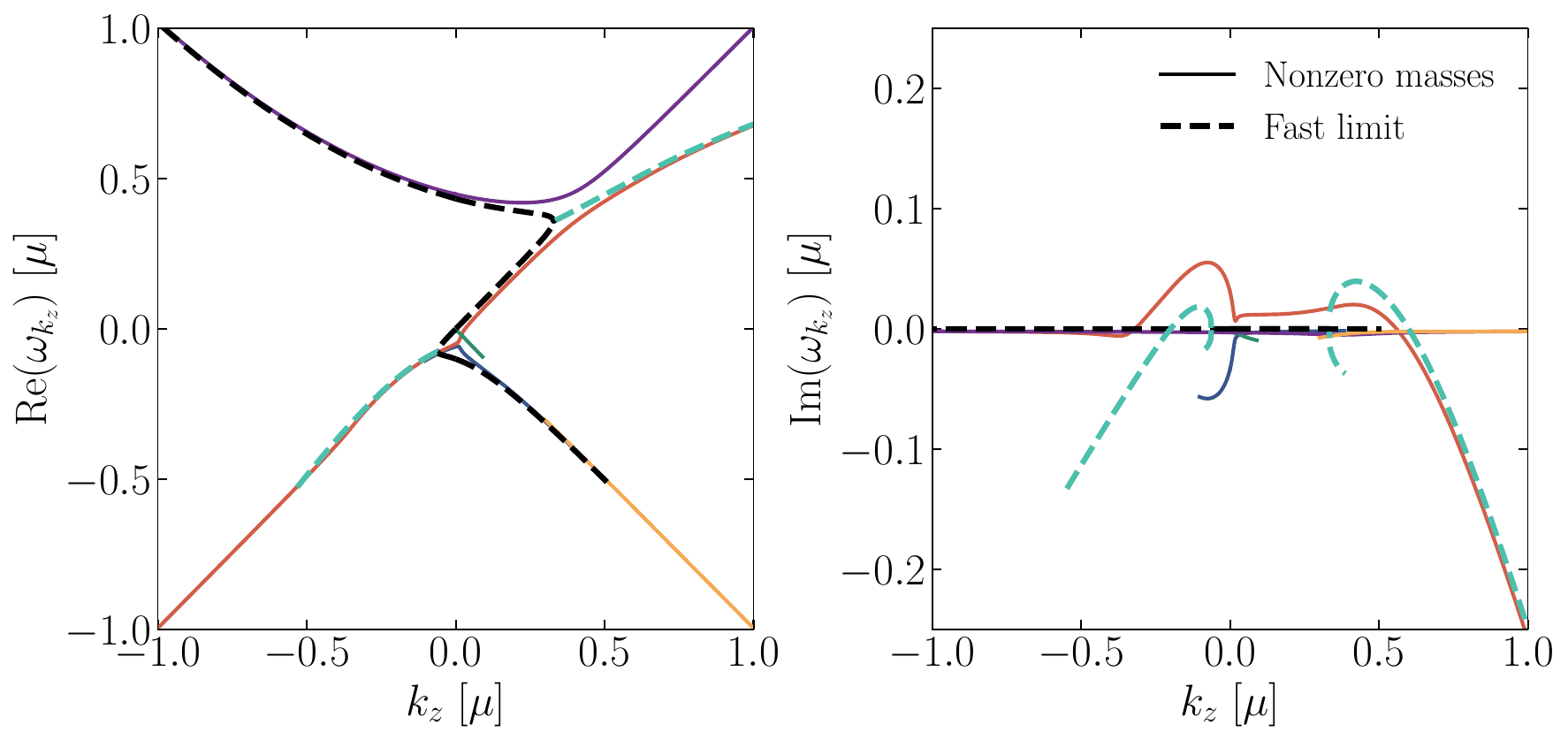}
    \caption{Real (left) and imaginary (right) part of the eigenfrequency as a function of wavevector along the symmetry axis $k_z$ for the benchmark case of a fast instability. The fast limit of vanishing neutrino masses is shown with thick dashed curves, in black for stable branches, and in teal for unstable and Landau-damped branches. For the case of nonzero masses, we show with different colors (blue, red, green, purple, orange) the five different branches found through numerical solution. The blue, green, and orange lines abruptly terminate when they touch a singular branch cut of the dispersion relation.}\label{fig:fast_splitting}
\end{figure*}

We now turn to another example for a case exhibiting a fast instability, i.e.\ an instability surviving in the limit of vanishing neutrino masses. The numerical solution for these cases, to our knowledge, has been usually given only neglecting neutrino masses. We now provide an explicit example where we turn on the neutrino mass splitting to consider its effect on the singularities of the flavor-wave dispersion.

In practice, we merely modify the benchmark DLN distribution of the previous example by endowing antineutrinos with an anisotropy measured by a parameter $\eta$

\begin{equation}
    D_{v,w}=\frac{1}{2\pi}\left[\frac{\Gamma_\nu n_\nu}{(w-w_\nu)^2+\Gamma_\nu^2}-\frac{\Gamma_{\overline{\nu}}n_{\overline{\nu}}(1+\eta v)}{(w-w_{\overline{\nu}})^2+\Gamma_{\overline{\nu}}^2}\right].
\end{equation}
The total DLN  is not affected, since the integral of $v$ from $-1$ to $+1$ vanishes. However, if $\eta$ is large enough, the total DLN at $v=1$ turns negative, creating an angular crossing and leading to a fast instability~\cite{Morinaga:2021vmc,Dasgupta:2021gfs,Fiorillo:2024bzm,Fiorillo:2024uki}. For our numerical example, we choose $\eta=1.2$. Due to the anisotropy, the frequency $\omega_{k_z}$ depends not just on the module but also on the sign of the wavevector along the symmetry axis $k_z$.

Figure~\ref{fig:fast_splitting} shows the solutions of the dispersion relation, comparing the fast limit of zero neutrino masses (thick, dashed lines) with the full solution for nonzero masses (solid lines). In the fast limit, the pattern is analogous to that discussed multiple times in the literature (e.g.\ Refs.~\cite{Yi:2019hrp,Fiorillo:2024dik,Kost:2026jrk}): a superluminal stable branch, shown in black, approaches the positive (negative) light cone at $k_z\to-\infty$ ($k_z\to+\infty$). The stable branch acquires a vertical tangent at two degeneracy points: at each one of them, it converts into a pair of complex conjugate solutions, shown in teal. The damped one---we emphasize that it is not Landau-damped, since it appears outside of the light cone---disappears when it touches the light cone, while the unstable one continues through the light cone until it becomes Landau-damped. The main feature we emphasize, in this fast limit, is the presence of exact degeneracy points, in agreement with our argument in Sec.~\ref{sec:degeneracy_superluminal} that, for a real-symmetric dispersion relation, such points are generically expected for a single parameter $k_z$.

\begin{figure*}
    \includegraphics[width=\textwidth]{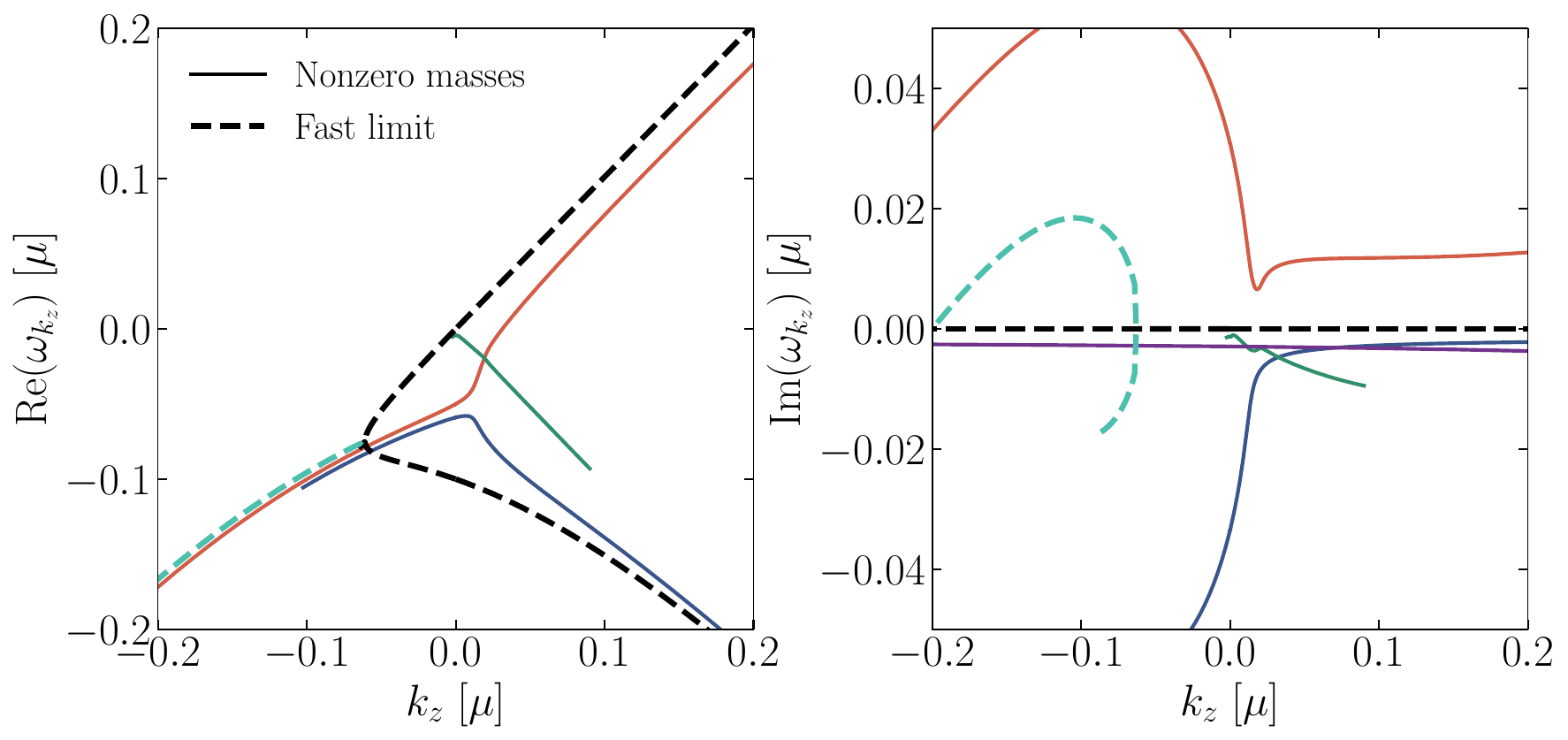}
    \caption{Zoomed-in version of Fig.~\ref{fig:fast_splitting}.}\label{fig:fast_splitting_zoom}
\end{figure*}

The nonzero neutrino mass splitting opens the gaps and removes the degeneracies. In turn, this leads to the appearance of multiple branches: we find overall five of them, shown with five different colors. Since some of them change very rapidly close to $k_z=0$, we also show a zoomed-in version of the figure for small $k_z$ in Fig.~\ref{fig:fast_splitting_zoom}.

Close to $k_z=0$, a green gapless band appears and disappears as it touches the branch cuts of the dispersion relation. In addition, the blue and red branches are the direct daughters of the stable branches shown in black from the fast limit. In particular, the two unstable branches for negative and positive $k_z$, which in the fast case were shown in teal and were separated by two degeneracy points, now form a single, smooth red branch continuing for all $k_z$, due to the removal of all degeneracies.

\section{Summary and discussion}\label{sec:summary}

The goal of this work was to determine what kinds of singularities can the flavor-wave frequency $\omega_\bk$ exhibit as a function of the wavevector $\bk$. There are several motivations behind this question. One is the generic ambition of understanding the properties of the flavor-wave dispersion intuitively: the dispersion relation is so complicated that it can only be solved numerically, and root finders easily show singular features that are easy to misinterpret as numerical artifacts. The only way to convince oneself of the reality of such features is to first predict them theoretically.

The second motivation is the recent realization that flavor waves are physical, emergent degrees of freedom that evolve together with neutrinos in the saturation of flavor instabilities~\cite{Fiorillo:2024qbl,Fiorillo:2024bzm,Fiorillo:2025npi,Fiorillo:2026byh,Fiorillo:2026tee}. Since the wavevector of these flavor waves changes with time, an immediate question is how do they behave as they cross singular points where their frequency changes suddenly; such question has been stressed especially in the formal treatment of Ref.~\cite{Johns:2025yxa} and in the numerical tests of Ref.~\cite{Kost:2026ckc}. It is essential here to take the continuous limit for the neutrino energy distribution, since for a few discrete neutrino energy modes and directions the nature of collective modes is entirely different---normal modes include the so-called Case-Van Kampen modes~\cite{Fiorillo:2024bzm} which disappear from the retarded response of the medium. While our analytical study of singularities cannot of course determine their role in the physical evolution of flavor waves, we believe such a clarification may play a basic role in that direction.

We show that, in the fast limit, degeneracies appear not at isolated points, but rather at surfaces in $\bk$ space. Previous studies focusing on axially symmetric waves only considered a single axis of symmetry, that intersects such surfaces along points. Furthermore, when nonzero neutrino masses are included, the degeneracy is generically lifted, and only one-dimensional lines of degeneracy are expected in $\bk$ space. Our numerical examples confirm this conclusion. Both for a slow and for a fast instability, we find that degeneracy points along the $z$ axis are lifted.

Termination singularities, namely modes abruptly terminating as they coincide with a branch cut of the flavor dielectric function, are generic and depend on the neutrino frequency and angular distribution. We have provided general rules that relate this distribution with the location of termination singularities by identifying their pinch singularities. Our numerical results completely confirm these rules.

As a practical byproduct, we have shown that representing the frequnecy distribution as a combination of Lorentzian distributions is an extremely useful tool for the solution of the dispersion relation. Each Lorentzian distribution behaves, from the perspective of the retarded response of the neutrino plasma, exactly like a monochromatic neutrino energy mode with an effective collision rate equal to the width of the distribution. The solution of the dispersion relation is thus significantly simplified in practice. As we have seen, using distributions that differ even very little from the true one close to the peak, may lead to entirely different results for the strongly damped modes with large negative $\mathrm{Im}(\omega_\bk)$. On the other hand, for precisely the same reason, these modes have little physical content---their fast damping makes them irrelevant in practice. In any case, it is clear that the tail of the frequency distribution, in a realistic astrophysical context, cannot be known with great precision anyway.

More importantly, while the singularities in our benchmark examples are of course determined by the use of Lorentzian distributions, we expect them to survive even for physical distributions. An easy argument leading to this conclusion is to note that the frequency distribution, when physical neutrino and antineutrino distributions are considered, must necessarily possess a discontinuity line at $\mathrm{Re}(w)=0$. The reason is that for $\mathrm{Re}(w)\to 0^+$ the frequency distribution is determined by the neutrino energy distribution, while for $\mathrm{Re}(w)\to 0^-$ it is determined by the antineutrino energy distribution, causing an unavoidable discontinuity---although exactly at $w=0$ the two must both vanish since there are no particles with infinite energy. The discontinuous behavior at $\mathrm{Re}(w)\to 0$ ensures the presence of singularities in the dispersion relation of flavor waves.

\acknowledgments

I thank Georg Raffelt for several useful conversations and comments on this manuscript, and Huaiyu Duan and Anson Kost for useful conversations on the topic. I also acknowledge support from the TAsP (Theoretical Astroparticle Physics) project.

\bibliographystyle{JHEP}
\bibliography{Biblio.bib}

\end{document}